\documentclass[aip,jcp,
reprint,
superscriptaddress,floatfix]{revtex4-1}

\usepackage[utf8]{inputenc}
\usepackage{color}
\usepackage{mathtools}
\usepackage{graphicx}
\usepackage{multirow}
\usepackage{amsmath,amssymb}

\usepackage{natmove}
\usepackage{braket}
\usepackage{bm}

\makeatletter
\newif\ifpreprintoption
\@ifclasswith{revtex4-1}{preprint}{\preprintoptiontrue}{\preprintoptionfalse}
\makeatother
\newcommand{\ifpreprint}[2]{\ifpreprintoption #1\else #2\fi}

\begin{document}

\title[]{Approximate analytical gradients and nonadiabatic couplings for the state-average density matrix renormalization group self-consistent field method}

\author{Leon Freitag}
\thanks{leon.freitag@phys.chem.ethz.ch (corresponding author)}
\affiliation{%
ETH Z\"urich, Laboratorium f\"ur Physikalische Chemie, Vladimir-Prelog-Weg~2, 8093 Z\"urich, Switzerland
}%
\author{Yingjin Ma}
% \affiliation{%
% ETH Z\"urich, Laboratorium f\"ur Physikalische Chemie, Vladimir-Prelog-Weg~2, 8093 Z\"urich, Switzerland
% }%
\affiliation{
Department of High Performance Computing, Computer Network Information Center, Chinese Academy of Sciences, Beijing 100190, China
}
\affiliation{
Center of Scientific Computing Applications \& Research, Chinese Academy of Sciences, Beijing 100190, China
}%
\author{Alberto Baiardi}
\affiliation{%
ETH Z\"urich, Laboratorium f\"ur Physikalische Chemie, Vladimir-Prelog-Weg~2, 8093 Z\"urich, Switzerland
}%
\author{Stefan Knecht}
\affiliation{%
ETH Z\"urich, Laboratorium f\"ur Physikalische Chemie, Vladimir-Prelog-Weg~2, 8093 Z\"urich, Switzerland
}%
\author{Markus Reiher}
\thanks{markus.reiher@phys.chem.ethz.ch (corresponding author)}
\affiliation{%
ETH Z\"urich, Laboratorium f\"ur Physikalische Chemie, Vladimir-Prelog-Weg~2, 8093 Z\"urich, Switzerland
}%

\date{September 28, 2019 (incl. minor proof corrections; Nov. 2019)}

\begin{abstract}

We present an approximate scheme for analytical gradients and nonadiabatic couplings for calculating state-average density matrix renormalization group self-consistent-field wavefunction. Our formalism follows closely the state-average complete active space self-consistent-field (SA-CASSCF) \emph{ansatz}, which employs a Lagrangian, and the corresponding Lagrange multipliers are obtained from a solution of the coupled-perturbed CASSCF (CP-CASSCF) equations. We introduce a definition of the matrix product state (MPS) Lagrange multipliers based on a single-site tensor in a mixed-canonical form of the MPS, such that a  sweep procedure is avoided in the solution of the CP-CASSCF equations. We apply our implementation to the optimization of a conical intersection in 1,2-dioxetanone, where we are able to fully reproduce the SA-CASSCF result up to arbitrary accuracy.

\end{abstract}

\maketitle

\section{Introduction}

Multiconfigurational methods are tailored for molecular systems exhibiting strong electronic correlation, as found in many transition metal complexes\cite{Pierloot_Calculations_TaCC_2005,Neese_Advanced_CCR_2007,Gagliardi_Transition_RiCC_2007,Neese_Correlated_JBIC_2011,Pierloot_Transition_IJQC_2011,Daniel_Photochemistry_CCR_2015,
Ashley_Ironing_CCR_2017,Radon_Benchmarking_PCCP_2019,Vogiatzis_Computational_CR_2019},
bond dissociation processes and excited electronic states.\cite{Gonzalez_Progress_C_2012,Plasser_Electronically_TCA_2012,Daniel_Photochemistry_CCR_2015,Ghosh_Combining_CR_2018,Lischka_Multireference_CR_2018} Excited electronic states are of key importance for photoinduced phenomena, including light-matter interaction with DNA\cite{Serrano-Andres_Are_JPPCPR_2009,Middleton_DNA_ARPC_2009,Mai_Excitation_2014}, light harvesting, photocatalysis and artificial photosynthesis\cite{Gust_Solar_ACR_2009,Gust_Realizing_FD_2012,Jager_Using_CCR_2015,Agbe_Recent_JIEC_2019}, as well as photodynamic cancer therapy\cite{Dolmans_Photodynamic_NRC_2003,Rose_Fiat_COCB_2008,Ormond_Dye_M_2013,Mehraban_Developments_M_2015}.

The majority of modern multiconfigurational methods are based on
the complete active space self-consistent field (CASSCF) method\cite{roos80,sieg81,shep87,olse11}, which, however, scales exponentially with the number of active orbitals, and as such, has been effectively limited to about 18 active orbitals.\cite{aqui15a} Only recently, this limit has been raised to 20 orbitals through massive parallelization.\cite{Vogiatzis_Pushing_JCP_2017} One successful remedy to the exponential scaling problem of CASSCF has been the
quantum chemical density matrix renormalization group (DMRG)\cite{White1998Ab, mitrushenkov2001quantum, chan2002highly, legeza2003controlling, chan2004algorithm, moritz2005convergence, moritz2005relativistic, rissler2006measuring, mcculloch2007density, lege08, chan2009density, mart10, schollwock2011density, chan2011density, Kurashige2009High,Knecht2014Communication, wout14, Yanai2015Density, knec16a, Chan2016Matrix}.
DMRG was originally introduced by White\cite{white1992density, white1992real} to solid state physics in 1992, but has found numerous applications in quantum chemistry since then.\cite{White_initio_JCP_1999,Daul_Full-CI_IJQC_2000,Mitrushenkov_Quantum_JCP_2001,Mitrushenkov_Quantum_JCP_2001,Mitrushenkov_Quantum_JCP_2003,Mitrushchenkov_importance_IJQC_2011,Chan_Highly_JCP_2002,Chan_Exact_JCP_2003,Chan_algorithm_JCP_2004,Chan_State-of-the-art_JCP_2004,Chan_Density-matrix_JCP_2005,Legeza_QC-DMRG_MP_2003,Moritz_Convergence_JCP_2005,Moritz_Relativistic_JCP_2005,hachmann2007radical,Zgid_spin_JCP_2008,Zgid_spin_JCP_2008,Zgid_density_JCP_2008,Zgid_Obtaining_JCP_2008,Kurashige_High-performance_JCP_2009}.
Combined with self-consistent-field orbital optimization (DMRG-SCF)\cite{zgid2008density, ghosh2008orbital, sun2017, wouters2014communication, Ma_Second-Order_JCTC_2017}, the method is able to approximate a CASSCF wave function to arbitrary accuracy with polynomial scaling\cite{chan2002highly} by reducing the size of the configurational space by means of the optimization of a so-called matrix product state (MPS) wave function\cite{Ostlund_Thermodynamic_PRL_1995,Rommer_Class_PRB_1997}, therefore allowing for much larger active spaces than traditional CASSCF.

Many problems in quantum chemistry tackled by DMRG and DMRG-SCF\cite{Chan_Chapter_ARiCC_2009,
mizukami2012more, kurashige2013entangled,Chalupsky_JAmChemSoc_Reactivity_2014,sharma2014low,yana15,freitag2015orbital, Olivaresamaya2015The,Wouters_density_EPJD_2014} require only the calculation of electronic energies and properties.
However, the study of photochemical phenomena often relies on optimizing excited-state structures, locating conical intersections and potential energy surface crossings, or  performing ab-initio molecular dynamic simulations. These tasks require efficient calculation not only of the energy of the ground and excited electronic states, but also of energy gradients and nonadiabatic couplings.

The most efficient way to calculate gradients is through the derivation of analytical expressions, as originally introduced by Pulay\cite{Pulay_initio_MP_1969,Pulay_Direct_AoEST_1977,Pulay_Systematic_JACS_1979,Pulay_efficient_TCA_1979}.
Accordingly, an analytical formulation for nonadiabatic couplings has been developed by Yarkony and co-workers.\cite{Iii1984On}

Analytical gradients are easily evaluated for a fully variationally-optimized wavefunction with the help of the Hellmann-Feynman theorem.\cite{Hellmann,Feynman_Forces_PR_1939} Single-state multiconfigurational self-consistent-field (MCSCF) wavefunctions are fully variationally optimized, and hence the state-specific CASSCF analytical gradient formulation \cite{Taylor_Analytical_JCC_1984} has appeared shortly after CASSCF has been introduced. The same holds true for single-state DMRG-SCF gradients, which have been introduced by \citet{Liu2013Multireference} and \citet{Hu2015Excited} and have found several applications in ground and excited state structure optimizations\cite{Liu2013Multireference, Hu2015Excited,Nakatani_JChemPhys_Density_2017} and resonance Raman spectra\cite{Ma2017Multiconfigurational}.

However, state-specific MCSCF methods have disadvantages, especially when applied to excited states. First, a specific state may not be tracked easily. For example, the so-called \emph{root flipping}, i.\,e. a change of the excited state order during orbital optimization may occur (although the problem can be partially alleviated if the state with the maximum overlap to the state of interest is followed during the optimization)\cite{Hu2015Excited,Ma2017Multiconfigurational}. 
Second, orthogonality, which greatly simplifies the calculation
of excited-state properties and, in particular, of nonadiabatic couplings\cite{Iii1984On,Yarkony_Electronic_1995,Schmidt_AnnuRevPhysChem_construction_2003} is not guaranteed in a state-specific optimization.

One possible remedy for the disadvantages of the state-specific optimization is the state-average MCSCF (SA-MCSCF) \emph{ansatz}\cite{Docken_JChemPhys_LiH_1972, Werner_JChemPhys_quadratically_1981, Diffenderfer_JPhysChem_Use_1982}, where several states are optimized simultaneously using a single set of molecular orbitals (MOs) which yields the best average energy of the states of interest. In the state-average \emph{ansatz}, the root flipping problem does not occur\cite{Docken_JChemPhys_LiH_1972} and the states are necessarily orthogonal.
However, with SA-MCSCF the gradient formalism loses its simplicity as the wavefunction is no longer fully variational. Nevertheless, the Hellmann-Feynman theorem can still be applied with the help of Lagrange multipliers\cite{Helgaker_Configuration-interaction_TCA_1989,Jones2001Analytical} that are obtained from the solution of the so-called coupled perturbed multiconfiguration self-consistent field (CP-MCSCF) equations.\cite{Dupuis_Energy_JCP_1981,Osamura_Generalization_JCP_1982}
\citet{Jones2001Analytical} were the first to describe an SA-MCSCF analytic gradient formulation based on Lagrangians. Recently, \citeauthor{Snyder_atomic_JCP_2015}\cite{Snyder_atomic_JCP_2015,Snyder_direct-compatible_JCP_2017} presented a GPU-based implementation of SA-MCSCF analytic gradients, following the work of \citeauthor{Jones2001Analytical}

Besides the gradients, the solution of the CP-MCSCF equations may also be used to calculate analytical nonadiabatic couplings in the SA-MCSCF formalism\cite{Iii1984On,Lengsfield_Nonadiabatic_ACP_2007} and for the second-order MCSCF orbital optimization procedure.\cite{wern80,wern85,know85,wern87,leng81,leng82,jorg81,yeag82,Olsen_Optimization_AiCP_1983}

While the SA-CASSCF analytical gradient problem may be considered solved, this is not the case for SA-DMRG-SCF analytical gradients.
The main challenge that remains is an adequate definition of the Lagrange parameters for the gradient calculation. For traditional MCSCF wavefunctions one usually expresses these in the configuration basis, which is impractical for MPS wavefunctions optimized by DMRG due to the exponential growth of the number of configurations with the number of active orbitals. The straightforward approach of employing all matrices in an MPS as parameters would result in a highly redundant parameterization, not least due to the gauge freedom of an MPS.\cite{Wouters_Thouless_PRB_2013,Nakatani_Linear_JCP_2014}. Nonredundant parametrizations of an MPS exist\cite{Dorando_Analytic_JCP_2009,Haegeman_Time-Dependent_PRL_2011,Wouters_Thouless_PRB_2013,Nakatani_Linear_JCP_2014,Vanderstraeten_Tangent-space_SPLN_2019}, the majority being based on the tangent-space concept.
However for large active spaces (which are often found in typical applications of SA-DMRG-SCF in quantum chemistry) the number of MPS parameters, even if they are non-redundant, would still become very large, possibly rendering the gradient calculation prohibitively expensive. Therefore, it is helpful to introduce an approximate parametrization, which does not necessarily cover the full MPS parameter space, but nevertheless allows for fast and efficient estimation of gradients.
In our previous work on a second-order DMRG-SCF optimization scheme\cite{Ma_Second-Order_JCTC_2017} we derived the expression of the state-average energy Hessian
based on an \textit{ansatz} for the MPS variational parameters. In this work, we generalize this \textit{ansatz} and employ it in a fast and efficient implementation to provide an efficient approximation for SA-DMRG-SCF analytical gradients and nonadiabatic couplings.

\section{Theory}\label{sec:method}
As the SA-DMRG-SCF analytical gradient theory is closely related to the SA-CASSCF analytical gradient theory, we will begin with a brief recap of the latter. Further information can be found in Ref.~\citenum{Jones2001Analytical} and especially the excellent paper by \citet{Snyder_direct-compatible_JCP_2017}.
In particular, we adopt the notation of the latter work. The theory for the nonadiabatic couplings follows very closely the gradient theory -- the differences between the two will be presented in Sec.~\ref{sec:nadc}.

\subsection{Single-state CASSCF gradients}

In single-state (or state-specific) CASSCF\cite{roos1980complete}, a wavefunction is defined as a linear combination of \emph{configurations} $\ket{\phi_I}$:
\begin{equation}
  \ket{\Phi} = \sum_I c_I\ket{\phi_I}\label{eqn:mc}
\end{equation}
where the expansion coefficients $c_I$ are called the configuration interaction (CI) coefficients and the configurations $\phi_I$ are chosen in such a way that they represent all possible excitations within a predefined active orbital space and do not include any excitations outside this space. The CI coefficients are obtained by  diagonalizing the matrix of the non-relativistic electronic Hamiltonian
  \begin{equation}
    \hat{H} = \sum_{\underset{\tau}{t,u}} \braket{t|h|u}a^\dagger_{t\tau}a_{u\tau} + \frac{1}{2}\!\!\sum_{\underset{\tau,\tau'}{t,u,v,w}}\!\! (tu|vw)a^\dagger_{t\tau}a^\dagger_{v\tau'}a_{w\tau'}a_{u\tau} + E_\text{core} \label{eqn:h}
  \end{equation}
  in the basis of the configurations $\phi_I$. Here, $\braket{t|h|u}$ and $(tu|vw)$ are the one- and two-electron integrals in a given orthonormal molecular orbital basis, respectively, and $a^\dagger_{t\tau}$ and $a_{t\tau}$ are creation and annihilation operators, respectively, for an orbital $t$ and spin $\tau$. $E_\text{core}$ is the sum of the energy contribution from the inactive (doubly occupied) orbitals and the nuclear repulsion energy.

The wavefunction $\ket{\Phi}$ is a function of a set of variational parameters, namely the orbital rotation parameters $\mathbf{\bm \kappa}$ and CI coefficients ${\bm c}$
\begin{equation}
  \Ket{\Phi} = \Phi(\mathbf{\bm \kappa},{\bm c})
\end{equation}
and is variationally optimized with respect to these parameters to yield a minimum single-state CASSCF energy
\begin{equation}
E^\Phi = \braket{\Phi|\hat{H}|\Phi}  = \sum_{pq}\Braket{p|h|q} \gamma_{pq} + \frac{1}{2} \sum_{pqrs} (pq|rs) \Gamma_{pqrs},\label{eqn:ss-casscf}
\end{equation}
with $\gamma_{pq}$ and $\Gamma_{pqrs}$ being the one- and two-particle reduced density matrices (RDMs), respectively. Since the optimized wavefunction $\ket{\Phi}$ is then fully variational with respect to all of its variational parameters, the energy gradient with respect to some perturbation $x$ (e.\,g.\ nuclear displacement) may be calculated according to the the Hellmann-Feynman theorem\cite{szalay2011multiconfiguration}:

\begin{align}
  \frac{dE^\Phi}{dx} &= \Braket{\Phi|\frac{\partial \hat{H}}{\partial x}|\Phi} \nonumber\\
  &= \sum_{pq}\frac{\partial\Braket{p|h|q}}{\partial x}\gamma_{pq} + \frac{1}{2} \sum_{pqrs} \frac{\partial (pq|rs)}{\partial x}\Gamma_{pqrs}.\label{eqn:ss-gradient}
\end{align}

\subsection{SA-CASSCF analytical gradients}

In SA-CASSCF, the wavefunction is determined by a variational optimization of the state-average energy of several states $\Psi$
\begin{equation}
  E^\text{SA} = \sum_\Psi \omega_\Psi E^\Psi = \sum_\Psi \omega_\Psi \braket{\Psi|\hat{H}|\Psi}\label{eqn:sa}
\end{equation}
where $\omega_\Psi$ is the weight of state $\Psi$, and the sum of all weights equals to 1. In this paper we will only consider a situation where all weights are equal,
although \citet{Jones2001Analytical} as well as \citet{Snyder_direct-compatible_JCP_2017} have also considered a non-equal-weights situation.
Similarly to the single-state case, the state-average energy depends on a set of variational parameters
\begin{equation}
  E^\text{SA} = E^\text{SA}(\mathbf{\bm \kappa},{\bm c}^1,{\bm c}^2,\ldots,{\bm c}^n)\label{eqn:parameters}
\end{equation}
where $\mathbf{\bm \kappa}$ are the orbital parameters and ${\bm c}^\Psi$ are the CI coefficients for state $\Psi$.

The optimization procedure ensures that the state-average energy, but not the individual state-specific energies, is variational with respect to the orbital parameters and the CI parameters.
\begin{equation}
  \frac{\partial E^\text{SA}}{\partial \kappa_{pq}} = 0;\quad%
  \frac{\partial E^\text{SA}}{\partial c_{\Theta I}} = 0;
  \quad \frac{\partial E^\Theta}{\partial c_{\Phi I}}\Bigg|_{\Theta\neq \Phi} \neq 0;\quad \frac{\partial E^\Theta}{\partial \kappa_{pq}} \neq 0;
\end{equation}

Hence, the gradient cannot be calculated according to the Hellmann-Feynman theorem. Following the method of the Lagrange multipliers, we may construct a Lagrangian $L^\Theta$ for state $\Theta$ starting from the energy in Eq.~\eqref{eqn:ss-casscf}
\begin{equation}\label{eqn:lagrange}
L^\Theta
=  E^\Theta + \left( \sum_{pq} \bar{\kappa}_{pq}^\Theta  \frac{\partial E^\text{SA}}{\partial \kappa_{pq}} -0 \right) + \left( \sum_{\Psi I} \bar{c}_{\Psi I}^{\Theta}  \frac{\partial E^\text{SA}}{\partial c_{\Psi I}} -0 \right).
\end{equation}
where $\bar{\kappa}_{pq}^\Theta$ and $\bar{c}_{\Psi I}^{\Theta}$ are the Lagrange multipliers corresponding to the orbital and CI parts, respectively. $L^\Theta$ is variational with respect to all parameters and, therefore, the Hellmann-Feynman theorem applies. The Lagrange multipliers can be obtained by exploiting the fact that the Lagrangian must be variational with respect to all of its parameters:
\begin{equation}\label{eqn:lag-eq}
\frac{\partial L^\Theta}{\partial \bar{\kappa}_{pq}^\Theta} = \frac{\partial L^\Theta}{\partial \bar{c}_{\Psi I}^{\Theta}} = \frac{\partial L^\Theta}{\partial \kappa_{pq}} = \frac{\partial L^\Theta}{\partial {c}_{\Psi I}} = 0.
\end{equation}
The first two equalities in Eq.~(\ref{eqn:lag-eq}) are trivially fulfilled and are equivalent to the definition of the constraints. Hence, the Lagrange multipliers are obtained from the other two equalities, yielding the \emph{coupled perturbed} MCSCF (CP-MCSCF) equations:

\begin{align}\label{eqn:lagrange_1}
\frac{\partial L^\Theta}{\partial {\kappa}_{rs}} = &
\underbrace{\frac{\partial E^\Theta}{\partial \kappa_{rs}}}_{g_{rs}^\Theta} +
\sum_{pq} \bar{\kappa}_{pq}^\Theta
\underbrace{\frac{\partial^2 E^\text{SA}}{\partial \kappa_{pq} \partial \kappa_{rs}}}_{H^{\text{OO}}_{pq,rs}} +
\sum_{\Psi I} \bar{c}_{\Psi I}^{\Theta}
\underbrace{\frac{\partial^2 E^\text{SA} }{\partial {c}_{\Psi I}\partial\kappa_{rs}}}_{H^{\text{CO}}_{\Psi I,rs}}
 = 0, \\
\frac{\partial L^\Theta}{\partial {c}_{\Phi J}} = &
\sum_{pq} \bar{\kappa}_{pq}^\Theta
\underbrace{\frac{\partial^2 E^\text{SA}}{\partial \kappa_{pq} \partial {c}_{\Phi J}}}_{H^{\text{OC}}_{pq,\Phi J}} +
\sum_{\Psi I} \bar{c}_{\Phi I}^{\Theta}
\underbrace{\frac{\partial^2 E^\text{SA}}{\partial {c}_{\Psi I} \partial {c}_{\Phi J}}}_{H^\text{CC}_{\Psi I \Phi J}}
=  0 \label{eqn:lagrange_2}
\end{align}
or in matrix form
\begin{equation}\label{eqn:lagrange-mat}
  \left(
        \begin{array}{cc}
          \mathbf{H^\text{OO}} & \mathbf{H^\text{CO}} \\
          \mathbf{H^\text{OC}} & \mathbf{H^\text{CC}} \\
        \end{array}
  \right)
  \left(
        \begin{array}{c}
          \mathbf{\bar{\bm\kappa}} \\
          \mathbf{\bar{c}^\Theta}
        \end{array}
  \right)
  = -
  \left(
        \begin{array}{c}
          \mathbf{g^\Theta} \\
          0
        \end{array}
  \right)
\end{equation}
As can be seen from Eqs.~(\ref{eqn:lagrange_1}) and (\ref{eqn:lagrange_2}), $\mathbf{H}$ is the state-average Hessian matrix with its orbital-orbital ($\mathbf{H}^\text{OO}$), orbital-CI ($\mathbf{H}^\text{OC}=(\mathbf{H}^\text{CO})^T$) and CI-CI ($\mathbf{H}^\text{CC}$) components and $\mathbf{g^\Theta}$ is the state-specific orbital gradient for state $\Theta$.

After solving Eq.~(\ref{eqn:lagrange-mat}), we may construct the Lagrangian of Eq.~(\ref{eqn:lagrange}), and calculate the gradient as follows\cite{Snyder_atomic_JCP_2015,Snyder_direct-compatible_JCP_2017}:

\begin{align}
  \frac{dE^\Theta}{dx}&= \sum_{pq}\frac{\partial\Braket{p|h|q}}{\partial x}\gamma^{\Theta,\text{e}}_{pq} + \frac{1}{2} \sum_{pqrs} \frac{\partial (pq|rs)}{\partial x}\Gamma^{\Theta,\text{e}}_{pqrs}- \nonumber\\
  &- \sum_{pq} X^{\Theta,\text{e}}_{pq} \frac{\partial S_{pq}}{\partial x}\label{eqn:sa-gradient}
\end{align}
The first two terms in Eq.~(\ref{eqn:sa-gradient}) originate from the Hellmann-Feynman theorem. They seem identical to Eq.~(\ref{eqn:ss-gradient}), but the one- and two-particle reduced density matrices have been replaced by their \emph{effective} counterparts, $\boldsymbol{\gamma}^{\Theta,\text{e}}$  and $\boldsymbol{\Gamma}^{\Theta,\text{e}}$, that are given by\cite{Snyder_direct-compatible_JCP_2017}:

\begin{align}
  \boldsymbol{\gamma}^{\Theta,\text{e}} &= \boldsymbol{\gamma}^\Theta + \boldsymbol{\tilde{\gamma}} + \boldsymbol{\bar{\gamma}} \label{eqn:oneptdm} \\
  \boldsymbol{\Gamma}^{\Theta,\text{e}} &= \boldsymbol{\Gamma^}\Theta + \boldsymbol{\tilde{\Gamma}} + \boldsymbol{\bar{\Gamma}} ,\label{eqn:twoptdm}
\end{align}
where $\boldsymbol{\tilde{\gamma}}$ and $\boldsymbol{\bar{\gamma}}$, as well as $\boldsymbol{\tilde{\Gamma}}$ and $\boldsymbol{\bar{\Gamma}}$ are the orbital and the CI contributions to the RDM, respectively:
\begin{align}
  \tilde\gamma_{pq} &= \sum_\Psi \omega_\Psi \left( \sum_o \gamma^\Psi_{oq}\bar{\kappa}^\Theta_{op} - \gamma^\Psi_{po}\bar{\kappa}^\Theta_{qo} \right) \\
  \bar\gamma_{pq} &= \sum_\Psi 2\omega_\Psi \gamma^{\Psi\boldsymbol{\bar{c}^\Theta_\Psi}}_{pq} \label{eqn:oneeff}
\end{align}
\begin{align}
  \tilde\Gamma_{pqrs} = \sum_\Psi \omega_\Psi \Bigg( \sum_o \Gamma^\Psi_{oqrs}\bar{\kappa}^\Theta_{op} &+ \Gamma^\Psi_{pors}\bar{\kappa}^\Theta_{oq}+ \\ \nonumber
  + \Gamma^\Psi_{pqos}\bar{\kappa}^\Theta_{or} &+ \Gamma^\Psi_{pqro}\bar{\kappa}^\Theta_{os} \Bigg) \\
  \bar\Gamma_{pqrs} = \sum_\Psi 2\omega_\Psi \gamma^{\Psi\boldsymbol{\bar{c}^\Theta_\Psi}}_{pqrs}. \label{eqn:twoeff}
\end{align}
Here $\gamma^{\Psi\boldsymbol{\bar{c}^\Theta_\Psi}}_{pq}$ and $\Gamma^{\Psi\boldsymbol{\bar{c}^\Theta_\Psi}}_{pqrs}$ are matrix elements of the one- and two-particle transition density matrices, respectively, between state $\Psi$ and a state with the Lagrange parameters $\boldsymbol{\bar{c}^\Theta_\Psi}$ as the CI coefficients. The last term in Eq.~(\ref{eqn:sa-gradient}) is dubbed the ``connection'' term by \citet{Jones2001Analytical}, and is evaluated from the effective CI Lagrangian and the derivative of the MO overlap matrix.\cite{Snyder_atomic_JCP_2015,Snyder_direct-compatible_JCP_2017}

In practice, the CP-MCSCF equations~(\ref{eqn:lagrange-mat}) are not solved directly, since the cost of evaluating and storing the full Hessian is too large. Instead, iterative solvers such as the preconditioned conjugate gradient (PCG)\cite{Jones2001Analytical} or the direct inversion of the iterative subspace (DIIS)\cite{Snyder_direct-compatible_JCP_2017} are employed, which only require the computation of the product between the Hessian and a trial vector, i.\,e. $\mathbf{H^\text{OO}}\tilde{\bm\kappa}^\Theta, \mathbf{H^\text{CO}}\tilde{\bm\kappa}^\Theta, \mathbf{H^\text{CO}}\tilde{\bm c}^\Theta$ and $\mathbf{H^\text{CC}}\tilde{\bm c}^\Theta$, with $\tilde{\bm\kappa}^\Theta$ and $\tilde{\bm c}^\Theta$ as trial vectors, which become equal to the corresponding Lagrange multipliers upon convergence. Although these quantities must be recalculated each iteration (by contrast, the Hessian needs to be calculated only once), one must store only a trial vector of size $n_{\bm{\kappa}}+n_{\bm{c}}$, where $n_{\bm{\kappa}}$ is the number of orbital rotation parameters
and $n_{\bm{c}}$ the number of CI coefficients, instead of a full Hessian matrix. In addition, as will become evident in Section~\ref{sec:impl}, evaluating certain matrix-vector products offers further computational advantages compared to evaluating the corresponding Hessian matrix elements.

\subsection{DMRG and DMRG-SCF}

Most commonly, DMRG for quantum chemistry is formulated on the basis of an MPS wavefunction parametrization\cite{Ostlund_Thermodynamic_PRL_1995,Rommer_Class_PRB_1997},
\begin{align}
  \Ket{\Psi} &=\sum_\sigma \mathbf{M}^{\sigma_1} \mathbf{M}^{\sigma_2}\cdots \mathbf{M}^{\sigma_L}\Ket{\bm{\sigma}} \nonumber\\
             &=\sum_\sigma\sum_{a_1\ldots a_{L-1}} M^{\sigma_1}_{1a_1} M^{\sigma_2}_{a_1a_2}\cdots M^{\sigma_L}_{a_{L-1}1}\Ket{\bm{\sigma}}.\label{eqn:mps}
\end{align}
Hence, the CI coefficients in Eq.~(\ref{eqn:mc}) are encoded as a product of three-dimensional tensors $\mathbf{M}^{\sigma_l}$, $\ket{\bm{\sigma}} = \ket{\sigma_1,\ldots,\sigma_L}$ represents the \emph{occupation number vector} in analogy to the configurations in Eq.~(\ref{eqn:mc}), and $L$ is the number of active orbitals. We may occasionally call the quantities $M^{\sigma_l}_{a_{l-1}a_l}$ the elements of an \emph{MPS tensor} at site (orbital) $l$.
For the optimization, it is important to express the MPS in a \emph{mixed-canonical form}\cite{schollwock2011density} at an arbitrary site $l$:
\begin{align}
  \Ket{\Psi}&=\mkern-20mu\sum_{\sigma,a_1\ldots a_{L-1}}\mkern-20mu{A}^{\sigma_1}_{1a_1}\cdots {A}^{\sigma_{l-1}}_{a_{l-2}a_{l-1}}{M}^{\sigma_{l}}_{a_{l-1}a_l} {B}^{\sigma_{l+1}}_{a_la_{l+1}}\cdots {B}^{\sigma_L}_{a_{L-1}1}\Ket{\sigma}, \label{eqn:mps-mixed-can}
\end{align}
where the tensors with elements ${A}^{\sigma_{i}}_{a_{l-1}a_{i}}$ are left-normalized and tensors with elements ${B}^{\sigma_{i}}_{a_{l-1}a_{i}}$ are right-normalized, respectively \cite{Schollwock_density-matrix_PTRSA_2011}.

In contrast with the CASSCF/CI wavefunction in Eq.~(\ref{eqn:mc}), for which all the CI coefficients are determined in one step, DMRG optimizes MPS wavefunctions iteratively with one (or two adjacent, see below) $\mathbf{M}^{\sigma_l}$ MPS tensors at a time, while ensuring that maximum dimensions of these matrices do not exceed a certain value $m$, which is denoted as the \emph{bond dimension} or \emph{number of renormalized block states}. Due to this systematic dimension reduction, the exponential scaling of Eq.~(\ref{eqn:mc}) is reduced to polynomial scaling. In analogy to Eq.~(\ref{eqn:mps}), also operators may be represented as matrix product operators (MPOs):
\begin{equation}
  \hat{\mathcal{W}} = \sum_{\sigma,\sigma'}\sum_{b_1\ldots b_{L-1}} W^{\sigma_1\sigma'_1}_{1b_1}W^{\sigma_2\sigma'_2}_{b_1b_2}\cdots W^{\sigma_L\sigma'_L}_{b_{L-1}1} \ket{\boldsymbol{\sigma}}\bra{\boldsymbol{\sigma'}}. \label{eqn:mpo}
\end{equation}
(We shall not go into detail here on how to obtain the MPO representation for various operators: this has been described in detail in our previous work\cite{kell15a}.)

In that case, the optimization of a MPS wavefunction is formulated as the variational optimization of the entries of a single MPS tensor (or two adjacent MPS tensors) to minimize the expectation value of the energy $E^\Psi=\braket{\Psi|\hat{H}|\Psi}$, under the constraint that the wavefunction is normalized:
\begin{equation}
  \delta\left(\braket{\Psi|\hat{H}|\Psi}-\lambda(\braket{\Psi|\Psi}-1)\right) = 0 \label{eqn:var-opt}
\end{equation}
Inserting the MPO expression for the Hamiltonian $\hat{H}$ into the expression for the expectation value of the energy, we obtain
\onecolumngrid
\begin{align}
  \braket{\Psi|\hat{H}|\Psi} &= \mkern-13mu\sum_{\underset{\sigma\sigma'}{\underset{a_1, \ldots, a_{L-1}}{a'_1, \ldots, a'_{L-1}}}}\mkern-13mu \left(M^{\sigma_1}_{1a_1}\cdots M^{\sigma_L}_{a_{L-1}1}\right)^*\mkern-13mu\sum_{b_1, \ldots, b_{L-1}}\mkern-13mu W^{\sigma_1\sigma'_1}_{1b_1}\cdots W^{\sigma_L\sigma'_L}_{b_{L-1}1}\left({M}^{\sigma'_1}_{1a'_1}\cdots {M}^{\sigma'_L}_{a'_{L-1}1}\right) \nonumber\\
  &=\mkern-27mu\sum_{\underset{\sigma_L\sigma'_L}{a_{L-1}a'_{L-1}b_{L-1}}}\mkern-27mu M^{\sigma_L\dagger}_{1a_{L-1}}W^{\sigma_L\sigma'_L}_{b_{L-1}1}\Big(\ldots \sum_{\underset{\sigma_2\sigma'_2}{a_1a'_1b_1}}M^{\sigma_2\dagger}_{a_2a_1}W^{\sigma_2\sigma'_2}_{b_1b_2}\Big(\sum_{\sigma_1\sigma'_1}M^{\sigma_1\dagger}_{a_11}W^{\sigma_1\sigma'_1}_{1b_1}{M}^{\sigma'_1}_{1a'_1}\Big){M}^{\sigma'_2}_{a'_1a'_2}\cdots\Big){M}^{\sigma'_L}_{a'_{L-1}1}\label{eqn:expval}
\end{align}
\ifpreprint{}{\twocolumngrid}
where we have regrouped the multiplication by indices $\sigma_i,\sigma'_i$ in the last step. We may now define \emph{left boundaries} recursively as\cite{schollwock2011density}
%\ifpreprint{}{\clearpage}
\begin{align}
  \mathbb{L}^{b_0}_{a_0a'_0} & = 1 \\
  \mathbb{L}^{b_1}_{a_1a'_1} & = \sum_{\sigma_1\sigma'_1}M^{\sigma_1\dagger}_{a_11}W^{\sigma_1\sigma'_1}_{1b_1}{M}^{\sigma'_1}_{1a'_1}, \\
  \cdots\nonumber\\
  \mathbb{L}^{b_l}_{a_la'_l} & = \mkern-18mu\sum_{\underset{\sigma_l\sigma_l'}{a_{l-1}a'_{l-1}b_{l-1}}}\mkern-18mu{}M^{\sigma_l\dagger}_{a_la_{l-1}}W^{\sigma_l\sigma'_l}_{b_{l-1}b_l}\mathbb{L}^{b_{l-1}}_{a_{l-1}a'_{l-1}}{M}^{\sigma'_l}_{a'_la'_{l-1}},\label{eqn:boundary}
\end{align}
and analogously \emph{right boundaries}
\begin{equation}
    \mathbb{R}^{b_{l-1}}_{a'_{l-1}a_{l-1}} = \sum_{\underset{\sigma_l\sigma_l'}{a_la'_lb_{l}}}{M}^{\sigma'_l}_{a'_{l-1}a'_l}W^{\sigma_l\sigma'_l}_{b_{l-1}b_l}\mathbb{R}^{b_l}_{a'_la_l}M^{\sigma_l\dagger}_{a_la_{l-1}}.
\end{equation}

Assuming that the left and right boundaries have been constructed from left- and right-normalized tensors, respectively, inserting the definition of boundaries into Eq.~(\ref{eqn:var-opt}) yields an eigenvalue equation $\mathcal{H}v = \lambda v$, where $\mathcal{H}$ is the local Hamiltonian matrix at site $l$ with matrix elements
\begin{align}
  \mathcal{H}_{IJ} &= \mathcal{H}_{(a_{l-1}\sigma_la_l),(a'_{l-1}\sigma'_la'_l)} = \mathcal{H}^{\sigma_l\sigma'_l}_{a_{l-1}a_l a'_{l-1}a'_l} \nonumber\\
  &=\sum_{b_{l-1}b_l} \mathbb{L}^{b_{l-1}}_{a_{l-1}a'_{l-1}}W^{\sigma_l\sigma'_l}_{b_{l-1}b_l}\mathbb{R}^{b_l}_{a'_la_l}, \label{eqn:local-h}\\
\intertext{and}
  v_I &= M_{(a_{l-1}\sigma_la_l)} = M^{\sigma_l}_{a_{l-1}a_l}, \label{eqn:local-hv}
\end{align}
where we combined one set of indices $(a_{l-1}\sigma_la_l)$ into one composite index to form the matrix $\mathcal{H}$ and the vector $v$. After the lowest (or several lowest, see below) eigenvalue(s) $\lambda$ have been obtained, the corresponding eigenvector(s) are reshaped again into the MPS tensor $M^{\sigma_l}_{a_{l-1}a_l}$. A subsequent normalization and truncation procedure such as singular value decomposition (SVD) ensures that the maximum dimension of ${M}^{\sigma_{l}}_{a_{l-1}a_{l}}$ does not exceed $m$. A basis transformation finalizes the local optimization at site $l$, generating a new mixed-canonical form of the MPS at site $l+1$. The process (\emph{sweep}) is repeated until the final site $L$ is reached and then its direction is reversed.  In passing, we note that commonly two adjacent MPS tensors are optimised simultaneously (two-site DMRG)\cite{schollwock2011density}, which is, however, not an important aspect for the purpose of this work.
For further details on the DMRG optimization procedure we refer the reader to Refs.~\citenum{schollwock2011density,Schollwock_density-matrix_PTRSA_2011,kell15a}.

After optimization of the MPS wavefunction, we can easily obtain one- and two-particle RDMs as expectation values of operators within the MPS-MPO framework of DMRG.\cite{schollwock2011density,Schollwock_density-matrix_PTRSA_2011} This allows us to formulate a (SA-)DMRG-SCF procedure in analogy with the (SA-)CASSCF one, with the CI coefficient optimization step in CASSCF replaced by a DMRG procedure in DMRG-SCF\cite{zgid2008density,Ma_Second-Order_JCTC_2017}. Importantly, Eqs.~(\ref{eqn:ss-casscf}) and (\ref{eqn:sa}) remain the same.

\subsection{Definition of variational parameters for analytical gradients for SA-DMRG-SCF}\label{subsec:def-param}

The derivation of the state-average gradient for an MPS wavefunction by means of Eqs.~(\ref{eqn:lag-eq})-(\ref{eqn:lagrange-mat}) is far from trivial. Unlike the expansion of Eq.~(\ref{eqn:mc}), the CI coefficients are not explicitly available for a MPS wavefunction in Eq.~(\ref{eqn:mps}). Hence, one must define an analogous set of \emph{MPS parameters} which are equivalent to the CI parameters ${\bm c}^1,{\bm c}^2,\ldots,{\bm c}^n$ in Eq.~(\ref{eqn:parameters}).

As mentioned in the Introduction, in this work we would like to avoid considering the set with all non-redundant entries of the MPS as free variational parameters. Instead, we introduce a much simpler, albeit an approximative parametrization of an MPS in the mixed canonical form where only the entries of the tensor associated to the canonization site are included in the variational space. We will justify the choice of our parametrization below.

Let us define two auxiliary MPSs in orbital subspaces spanned by sites 1 to $l-1$ and $l+1$ to $L$:
\begin{align}
  \Ket{a_{l-1}} &= \sum_{\sigma_1\ldots\sigma_{l-1},a_1\ldots{}a_{l-2}}{A}^{\sigma_1}_{1a_1}\cdots {A}^{\sigma_{l-1}}_{a_{l-2}a_{l-1}} \ket{\sigma_1\ldots\sigma_{l-1}}\label{eqn:mpsa} \\
  \Ket{a_{l}} &= \sum_{\sigma_{l+1}\ldots\sigma_{L},a_{l+1}\ldots{}a_{L}} {B}^{\sigma_{l+1}}_{a_{l}a_{l+1}}\cdots {B}^{\sigma_L}_{a_{L-1}1}\Ket{\sigma_{l+1}\ldots\sigma_{L}}. \label{eqn:mpsb}
\end{align}
Inserting these equations into the mixed-canonical representation of our MPS at site $l$ (Eq.~(\ref{eqn:mps-mixed-can})) yields
\begin{equation}
  \Ket{\Psi} = \sum_{\sigma_l,a_{l-1},a_{l+1}} {M}^{\sigma_{l}}_{a_{l-1}a_{l}}\Ket{a_{l-1}}\otimes\ket{\sigma_{l}}\otimes\Ket{a_{l}} \label{eqn:mps-basis}
\end{equation}
where $\ket{\sigma_l}$ is the local basis state for orbital $l$, corresponding to its four possible occupations: unoccupied, spin-up, spin-down, and doubly occupied. Grouping the indices $(a_{l-1}\sigma_la_l)$ as in Eqs.~(\ref{eqn:local-h}) and (\ref{eqn:local-hv}), we arrive at an expression for $\ket{\Psi}$ equivalent to Eq.~(\ref{eqn:mc}):
\begin{equation}
  \ket{\Psi} = \sum_I v_I \ket{\Phi_I}\label{eqn:mps-compact}
\end{equation}
(From now on indices $I$ and $J$ in equations containing MPS will refer to the grouped indices $(a_{l-1}\sigma_la_l)$, in analogy to CI configurations, which are also labelled $I$ in Eq.~(\ref{eqn:mc}).

With a wavefunction written in terms of a CI expansion, we may define MPS Lagrange multipliers (or variational parameters)  $\bar{\bm v}^\Psi$ in a basis spanned by $v_I$ in analogy to the CI Lagrange multipliers $\bar{\bm c}^\Psi$ and construct a Lagrangian for a DMRG-SCF wavefunction similarly to Eq.~(\ref{eqn:lagrange}) employing MPS Lagrange multipliers.

  Such a definition of the variational parameters places a constraint on some properties of the MPS.
  The configuration basis $\ket{\phi_i}$ in Eq.~(\ref{eqn:mc}) is, by definition, the same for all states in the state-average description. This is not necessarily true for a set of MPS, if they are optimized individually because $\ket{a_{l-1}}$ and $\ket{a_l}$ (Eq.~\ref{eqn:mps-basis}) will be different for each state. Moreover, in this case not even the dimensions of the individual ${M}^{\sigma_{l}}_{a_{l-1}a_{l}}$ may be the same for all states. Consequently, for our definition of MPS parameters to be valid, the basis $\ket{\phi_i}$ in Eq.~(\ref{eqn:mps-compact}) must also be \emph{the same for all states in the state-average description}. This can be ensured if all MPS are optimized and brought into a mixed-canonical form simultaneously, where the truncation and the basis transformation step during the sweeps are the same for all states. In practice, this means employing \emph{common left- and right boundaries for all states} by following the procedure described in Ref.~\citenum{Dolgov2014_BlockEigenvalues}. We have implemented such a procedure in our new DMRG MPS-MPO solver: the details of the implementation are, however, beyond the scope of this work and will be described in a future publication.

  It remains to be specified at which site $l$ we write the MPS in Eq.~(\ref{eqn:mps-mixed-can}) to define our MPS parameters, as the choice is, in principle, arbitrary. However, the site must be chosen a priori before solving the CP-MCSCF equations. We will call this site the \emph{linear response site} for convenience. At first glance, there should be no difference, because for an optimized MPS wavefunction all mixed canonical forms at all possible linear response sites $l$ are equivalent. However, the parameters $v_I$ and the bases $\ket{\phi_I}$ for different linear response sites are not equivalent. In fact, here the approximative character of our MPS parametrization is revealed. We consider the variation of MPS parameters $v_I$ at a single site, but not the variation of the basis $\ket{\phi_I}$ that depends on the MPS tensors on the sites different from the canonization one. With sites towards the middle of the lattice, the number of parameters grows, and therefore the variational parameter space covered by the MPS parameters $v_I$ is enlarged. As the MPS approaches the full CI wavefunction, the variation of the basis $\ket{\Phi_I}$ approaches zero and the error arising from neglecting the variation of the basis (which we call \emph{basis-variation omission error (BVOE)}) vanishes, although, in general, for approximate full CI wavefunctions it is nonzero. We will assess the accuracy of the gradient calculation for different linear response sites on an example in Section~\ref{sec:cyclobutadiene}, and quantify BVOE in Section~\ref{sec:numerical}.

  In passing we note that in Ref.~\citenum{Ma_Second-Order_JCTC_2017} we have employed a similar definition of MPS parameters, where, in contrast to this work, i) two-site tensors at the first two sites have been employed in the definition instead of one-site MPS tensors at an arbitrary site and ii) the requirement of the same local basis $\ket{\Phi_i}$ for several states was not met. We consider the parametrization employed in this work as an improvement to the one from Ref.~\citenum{Ma_Second-Order_JCTC_2017}, as a response site in the middle of a lattice delivers larger variational flexibility of the wavefunction compared to the sites at the edge.

  \subsection{Implementation of SA-DMRG-SCF analytical gradients}\label{sec:impl}
  The definition of MPS parameters in Subsection~\ref{subsec:def-param} and the identification of a local optimization of an MPS tensor ${M}^{\sigma_{l}}_{a_{l-1}a_{l}}$ as CI problem allow us to formulate the SA-DMRG-SCF analytical gradient theory very closely following the SA-CASSCF theory, i.\,e. by applying Eqs.~(\ref{eqn:lag-eq})-(\ref{eqn:lagrange-mat}) to SA-DMRG-SCF wavefunctions (which have been optimized as described in Subsection~\ref{subsec:def-param}). This has another advantage, namely that we may derive our implementation from an existing SA-CASSCF analytical gradient one. In this work, we have based our implementation on the SA-CASSCF analytical gradient implementation in OpenMOLCAS.\cite{Bernhardsson_MolPhys_direct_1999,Jones2001Analytical,OpenMOLCAS} It employs the PCG method to solve the CP-MCSCF equations~(\ref{eqn:lagrange-mat}) iteratively, evaluating the Hessian-trial vector products in each iteration.

  Below we shall outline the quantities required for the analytical gradient calculation, whose evaluation is specific to the MPS wavefunction.

  \paragraph{Transition density matrices.} The evaluation of transition density matrix elements may be performed with the help of Eq.~(\ref{eqn:expval})\cite{kell15a}. For an MPS
    \begin{equation}
    \ket{\bar{\Psi}} = \sum_\sigma\sum_{a_1\ldots a_{L-1}} \bar{M}^{\sigma_1}_{1a_1} \bar{M}^{\sigma_2}_{a_1a_2}\cdots \bar{M}^{\sigma_L}_{a_{L-1}1}\Ket{\sigma} \label{eqn:mpstilde}
  \end{equation}
  an expectation value $\braket{\Psi|\hat{\mathcal{W}}|\bar{\Psi}}$ may be calculated as
  \onecolumngrid
    \begin{equation}
     \braket{\Psi|\hat{\mathcal{W}}|\bar{\Psi}} = \mkern-27mu\sum_{\underset{\sigma_L\sigma'_L}{a_{L-1}a'_{L-1}b_{L-1}}}\mkern-27mu M^{\sigma_L\dagger}_{1a_{L-1}}W^{\sigma_L\sigma'_L}_{b_{L-1}1}\Big(\ldots \sum_{\underset{\sigma_2\sigma'_2}{a_1a'_1b_1}}M^{\sigma_2\dagger}_{a_2a_1}W^{\sigma_2\sigma'_2}_{b_1b_2}\Big(\sum_{\sigma_1\sigma'_1}M^{\sigma_1\dagger}_{a_11}W^{\sigma_1\sigma'_1}_{1b_1}\bar{M}^{\sigma'_1}_{1a'_1}\Big)\bar{M}^{\sigma'_2}_{a'_1a'_2}\cdots\Big)\bar{M}^{\sigma'_L}_{a'_{L-1}1}\label{eqn:tdm}
  \end{equation}
  \ifpreprint{}{\twocolumngrid}
  To exploit this for the one- and two-particle reduced transition density matrix elements $\gamma^{\Psi\bar\Psi}_{pq}$ and $\Gamma^{\Psi\bar\Psi}_{pqrs}$, we replace $\hat{\mathcal{W}}$ in Eq.~(\ref{eqn:tdm}) with the MPO form of the operators

  \begin{align}
    \sum_\tau a^\dagger_{p\tau}a_{q\tau}& \qquad \text{for } \gamma^{\Psi\bar\Psi}_{pq}, \\
    \sum_{\tau\tau'} a^\dagger_{p\tau}a^\dagger_{r\tau'}a_{s\tau'}a_{q\tau}& \qquad \text{for } \Gamma^{\Psi\bar\Psi}_{pqrs}
  \end{align}
  respectively, and evaluate Eq.~(\ref{eqn:tdm}). (For details on obtaining the MPO form of various operators, see Ref.~\citenum{kell15a}).

  If the bra and ket in Eq.~(\ref{eqn:tdm}) differ only by a single tensor, then, instead of recalculating the full Eq.~(\ref{eqn:tdm}) for every expectation value one must calculate all contractions for all sites $\neq l$ only once for all expectation values and recalculate only the contractions involving tensors at site $l$.
  During the solution of the CP-MCSCF equations, only the tensor at site $l$ is updated. This can be seen in Eqs.~(41) and (72) in the work of \citet{Snyder_direct-compatible_JCP_2017}, who update the CI coefficients but retain the basis intact; by analogy, we update only the vector $v_I$ in Eq.~(\ref{eqn:mps-compact}) and therefore only the MPS tensor at site $l$. Furthermore, in the framework of SA-DMRG-SCF, $\ket{\bar{\Psi}}$ always differs from $\ket{\Psi}$ only by a single tensor, as multiple states are optimized with a common boundary. However, after a complete sweep the differing site is always at the beginning of the lattice. Hence, to bring the MPS into a usable form, we must perform a simultaneous canonization of all states. Details on this procedure are beyond the scope of this work and will be described in a future publication.

  \paragraph{Sigma vectors.} Calculation of Hessian-trial vector products requires calculation of sigma vectors. For a CI wavefunction Eq.~(\ref{eqn:mc}), a sigma vector is defined as the left-hand side of the CI eigenvalue equation, i.\,e.
  \begin{equation}
    \Sigma_i = \sum_j H_{ij} c_j
  \end{equation}
  where $H_{ij}$ is the matrix element of the Hamiltonian. (Note that in contrast to other literature, e.\,g.\ Refs.~\citenum{Jones2001Analytical}, \citenum{Snyder_direct-compatible_JCP_2017}, \citenum{helgaker2014molecular} we denote the sigma vector with a capital $\Sigma$ to avoid confusion with MPS indices $\sigma$). 
  An expression for the sigma vector for MPS wavefunctions is easily obtained from Eqs.~(\ref{eqn:local-h}) and (\ref{eqn:local-hv})
  \begin{align}
    \Sigma^{\sigma_l}_{a_{l-1}a_l} &= \sum_{a'_{l-1}a'_l\sigma'_l}\mathcal{H}^{\sigma_l\sigma'_l}_{a_{l-1}a_l a'_{l-1}a'_l}M^{\sigma'_l}_{a'_{l-1}a'_l} \nonumber\\
    &= \sum_{a'_{l-1}a'_l\sigma'_l}\sum_{b_{l-1}b_l} \mathbb{L}^{b_{l-1}}_{a_{l-1}a'_{l-1}}W^{\sigma_l\sigma'_l}_{b_{l-1}b_l}\mathbb{R}^{b_l}_{a'_la_l}M^{\sigma'_l}_{a'_{l-1}a'_l} \label{eqn:sigmavec}
  \end{align}
  Also in DMRG this equation constitutes the left-hand side of the local eigenvalue equation for the optimization of an MPS tensor, and, as such, belongs to the core part of our DMRG MPS-MPO implementation.\cite{kell15a}

  \paragraph{Diagonal of the Local Hamiltonian.} Also diagonal elements of the local Hamiltonian $\mathcal{H}^{\sigma_l\sigma'_l}_{a_{l-1}a_l a'_{l-1}a'_l}$ from Eq.~(\ref{eqn:local-h}) are required in our gradient implementation to calculate the preconditioner for the PCG algorithm.

  With the above quantities, we may setup the CP-MCSCF equations and evaluate the gradients. We will closely follow the work of \citet{Snyder_direct-compatible_JCP_2017} on SA-CASSCF gradients and present here only DMRG-specific steps in order to be as brief as possible.

  Equivalently to SA-CASSCF theory, we must set up and solve the CP-DMRG-SCF equations
  \begin{equation}\label{eqn:lagrange-mat-v}
    \left(
          \begin{array}{cc}
            \mathbf{H^\text{OO}} & \mathbf{H^\text{CO}} \\
            \mathbf{H^\text{OC}} & \mathbf{H^\text{CC}} \\
          \end{array}
    \right)
    \left(
          \begin{array}{c}
            \mathbf{\bar{\bm\kappa}} \\
            \bar{\bm v}^\Theta
          \end{array}
    \right)
    = -
    \left(
          \begin{array}{c}
            \mathbf{g^\Theta} \\
            0
          \end{array}
     \right)
  \end{equation}
  where the only difference to Eq.~(\ref{eqn:lagrange-mat}) is that the Lagrange CI parameters $\mathbf{\bar{c}^\Theta}$ have been replaced with the MPS parameters $\bar{\bm v}^\Theta$. As we employ the PCG solver\cite{Jones2001Analytical} for the CP-DMRG-SCF equations, we must be able to compute the Hessian-trial vector products  $\mathbf{H^\text{OO}}\tilde{\bm\kappa}^\Theta, \mathbf{H^\text{CO}}\tilde{\bm v}^\Theta,\mathbf{H^\text{CO}}\tilde{\bm\kappa}^\Theta$ and $\mathbf{H^\text{CC}}\tilde{\bm v}^\Theta$ with $\tilde{\bm\kappa}^\Theta$ and $\tilde{\bm v}^\Theta$ as trial vectors (which become equal to $\mathbf{\bar{\bm\kappa}}$ and $\bar{\bm v}^\Theta$ at convergence). $\mathbf{H^\text{OO}}\tilde{\bm\kappa}^\Theta$ and $\mathbf{g^\Theta}$ are evaluated exactly as for SA-CASSCF (see, e.\,g.,\ Eqs. (13)-(21) in \citet{Snyder_direct-compatible_JCP_2017}). To express the remaining Hessian-trial vector products, we introduce the generalized orbital gradient matrix $\mathbf{T}(\Psi,\bar{\Psi})$. This matrix has been introduced first in \citet{Snyder_direct-compatible_JCP_2017} (see Eqs.~(35)-(37) therein). However, in our implementation we followed Refs.~\citenum{Bernhardsson_MolPhys_direct_1999,Jones2001Analytical} and the standard textbook by Helgaker, J\o{}rgensen and Olsen\cite{helgaker2014molecular}. The equations for the calculation of the matrix elements $T_{pq}(\Psi,\bar{\Psi})$ as employed in our implementation are provided below:

  \begin{equation}
    T_{pq}(\Psi,\bar{\Psi}) = 2(F_{pq}(\Psi,\bar{\Psi})-F_{qp}(\Psi,\bar{\Psi})),\label{eqn:tmat}
  \end{equation}
  where the generalized Fock matrix $F_{pq}(\Psi,\bar{\Psi})$ is calculated as
  \begin{align}
    F_{ip}(\Psi,\bar{\Psi}) &= 2\left(^\text{I}F_{pi}\delta_{\Psi\bar{\Psi}}+^\text{A}F_{pi}(\Psi,\bar{\Psi})\right) \label{eqn:fock1}\\
    \intertext{if the first index is inactive and the second is arbitrary and as,}
    F_{tp}(\Psi,\bar{\Psi}) &= \sum_u {^\text{I}F}_{pu}\gamma_{tu}^{\Psi\bar\Psi} + Q_{tp} \nonumber\\
                            &= \sum_u {^\text{I}F}_{pu}\gamma_{tu}^{\Psi\bar\Psi} + \sum_{vwx} \Gamma_{tvwx}^{\Psi\bar\Psi}(pv|wx) \\
    \intertext{if the first index is active and the second is arbitrary, and}
    F_{ap}(\Psi,\bar{\Psi}) &= 0
  \end{align}
  if the first index is virtual. $^\text{I}F_{pq}$ and $^\text{A}F_{pq}(\Psi,\bar{\Psi})$ are the \emph{inactive} and \emph{active} Fock matrices, respectively:
  \begin{align}
    ^\text{I}F_{pq} &= \braket{p|h|q} + \sum_i \left(2(pq|ii)-(pi|iq)\right),\\
    ^\text{A}F_{pq}(\Psi,\bar{\Psi}) &= \sum_{vw} \gamma_{vw}^{\Psi\bar\Psi}\left( (pq|vw)-\frac{1}{2}(pw|vq)\right).\label{eqn:act-fock}
  \end{align}

  Eqs.~(\ref{eqn:fock1})-(\ref{eqn:act-fock}) are generalizations of Eqs.~(10.8.27)-(10.8.32) from Ref.~\citenum{helgaker2014molecular} to transition density matrices between states $\Psi$ and $\bar\Psi$. In particular, the state-specific orbital gradient is obtained with $g_{pq}^\Theta = T_{pq}(\Theta,\Theta)$.

  Similarly to \citet{Snyder_direct-compatible_JCP_2017}, the $\mathbf{H^\text{CO}}\tilde{\bm v}^\Theta$ product is evaluated as
  \begin{equation}
    \left(\sum_{\Psi I} H^\text{CO}_{\Psi I,pq}\tilde{v}^\Theta_{\Psi I}\right)_{pq} = \sum_\Psi 2\omega_\Psi T_{pq}(\Psi, \tilde{\bm v}_\Psi^\Theta),\label{eqn:HcoC}
  \end{equation}
  where $T_{pq}(\Psi, \tilde{\bm v}_\Psi^\Theta)$ indicates a matrix element of the $\mathbf{T}$ matrix calculated for a state $\Psi$, where the MPS tensor at the linear response site $l$ has been replaced by the corresponding block of the trial vector of MPS parameters $\tilde{\bm v}_\Psi^\Theta$, and an unmodified state $\Psi$.

  The $\mathbf{H^\text{CO}}\tilde{\bm\kappa}^\Theta$ product is evaluated from
  \begin{equation}
    \left(\sum_{pq} H^\text{CO}_{\Psi I,pq}\tilde{\kappa}^\Theta_{pq}\right)_{\Psi I} = 2\omega_\Psi\left(\tilde{\Sigma}_I^\Psi-\sum_\Phi v_I^\Phi R^{\Psi\Phi}\right)\label{eq:Hcc-c}
  \end{equation}
  with
  \begin{equation}
    R^{\Psi\Phi} = \sum_J \tilde{\Sigma}_J^\Psi v_J^\Phi.
  \end{equation}
  The modified sigma vector $\tilde{\Sigma}^\Psi_I$ is evaluated from Eq.~(\ref{eqn:sigmavec}), where, however, a modified Hamiltonian
  \begin{equation}
    \hat{\tilde{H}} = \sum_{\underset{\tau}{t,u}} \tilde{\braket{t|h|u}}a^\dagger_{p\tau}a_{q\tau} + \frac{1}{2} \sum_{\underset{\tau,\tau'}{t,u,v,w}} \tilde{(tu|vw)}a^\dagger_{t\tau}a^\dagger_{v\tau'}a_{w\tau'}a_{u\tau} \label{eqn:htilde}
  \end{equation}
  with transformed integrals\cite{helgaker2014molecular}

  \begin{equation}
    \tilde{\braket{t|h|u}} = \sum_p \tilde{\kappa}^\Theta_{tp} {}^\text{I}F_{pu} + \sum_{pq}\tilde{\kappa}^\Theta_{pq}\left(2(pq|tu)-(pt|qu)\right)
  \end{equation}%
  \begin{align}
  \intertext{and}
    \tilde{(tu|vw)} = \sum_p \Big( \tilde{\kappa}^\Theta_{tp}(pu|vw )&-\tilde{\kappa}^\Theta_{pu}(tp|vw) + \nonumber\\
                    +\tilde{\kappa}^\Theta_{vp}(tu|pw) &-\tilde{\kappa}^\Theta_{pw}(tu|vp) \Big)
  \end{align}
  is employed instead of the Hamiltonian in the form of Eq.~(\ref{eqn:h}).

  We should also note that calculating the $\mathbf{H^\text{CO}}\tilde{\bm\kappa}^\Theta$ product yields another computational advantage over the calculation of the full Hessian block $\mathbf{H^\text{CO}}$ (in addition to the obvious advantage of not storing the full Hessian). The calculation of $\mathbf{H^\text{CO}}$ would require one- and two-particle RDM derivatives with respect to MPS parameters\cite{wern80,Ma_Second-Order_JCTC_2017}, which we have shown in previous work\cite{Ma_Second-Order_JCTC_2017}, has the cost of an evaluation of a transition density matrix for each single MPS parameter. As the number of MPS parameters may become extremely large, albeit limited by $m$, the cost of evaluating RDM derivatives would easily become a bottleneck of the calculation. However, calculating $\mathbf{H^\text{CO}}\tilde{\bm\kappa}^\Theta$ by means of Eq.~(\ref{eq:Hcc-c}) avoids the calculation of RDM derivatives altogether.

  The $\mathbf{H^\text{CC}}\tilde{\bm v}^\Theta$ product, similarly to the CI case (cf.~Eq.~(39) in Ref.~\citenum{Snyder_direct-compatible_JCP_2017} or Eq.~(30) in Ref.~\citenum{Jones2001Analytical}), is evaluated from
  \begin{align}
    \left(\sum_{\Phi J} H^\text{CC}_{\Psi I,\Phi J}\tilde{v}^\Theta_{\Phi J}\right)_{\Psi I} &= 2\omega_\Psi\left(\sum_J \mathcal{H}_{IJ} \tilde{v}^\Theta_{\Psi J} - E^\Psi \tilde{v}^\Theta_{\Psi I}\right) \nonumber\\
    &= 2\omega_\Psi\left(\bar{\Sigma}^\Psi_I - E^\Psi \tilde{v}^\Theta_{\Psi I}\right).
  \end{align}
  Here we must again calculate a sigma vector $\bar{\Sigma}^\Psi_I$ with the help of Eq.~(\ref{eqn:sigmavec}), where, similarly to  Eq.~(\ref{eqn:HcoC}), the MPS tensor $\{M^{\sigma_l}_{a_{l-1}a_l}\}$ is replaced with the block of the trial vector of MPS parameters $\tilde{\bm v}^\Theta_{\Psi}$, reshaped according to the MPS tensor structure.

  After solving the CP-DMRG-SCF equations we obtain a set of the orbital and MPS parameters $\mathbf{\bar{\bm\kappa}}$ and $\bar{\bm v}^\Theta$, which allow us to construct the effective one- and two-particle RDMs from Eqs.~(\ref{eqn:oneptdm})-(\ref{eqn:twoeff}). The transition density matrices
  $\gamma^{\Psi\boldsymbol{\bar{v}^\Theta_\Psi}}_{pq}$ and $\Gamma^{\Psi\boldsymbol{\bar{v}^\Theta_\Psi}}_{pqrs}$, analogs of their CI counterparts $\gamma^{\Psi\boldsymbol{\bar{c}^\Theta_\Psi}}_{pq}$ and $\Gamma^{\Psi\boldsymbol{\bar{c}^\Theta_\Psi}}_{pqrs}$ from Eqs.~(\ref{eqn:oneeff}) and (\ref{eqn:twoeff}), are obtained as in Eq.~(\ref{eqn:HcoC}), namely by replacing the MPS tensor  $\{M^{\sigma_l}_{a_{l-1}a_l}\}$  in the MPS $\Psi$ at the linear response site $l$ with the vector of MPS parameters $\bar{\bm v}^\Theta_{\Psi}$, reshaped accordingly, and evaluating the transition density matrices between the unmodified $\Psi$ and the modified state. Finally, the effective density matrices allow us to construct the Lagrangian in Eq.~(\ref{eqn:lagrange}) and evaluate the gradient according to Eq.~(\ref{eqn:sa-gradient}). This is completely identical to the SA-CASSCF procedure and has been described extensively in Refs.~\citenum{Jones2001Analytical} and \citenum{Snyder_direct-compatible_JCP_2017}.

\subsection{Nonadiabatic couplings}\label{sec:nadc}
As in the SA-CASSCF case \cite{Fdez.Galvan_Analytical_JCTC_2016,Snyder_direct-compatible_JCP_2017}, the calculation of the nonadiabatic couplings for two states $\Theta$ and $\Lambda$, $\Braket{\Theta|{\partial}/{\partial x}| \Lambda}$, is very similar to the calculation of the gradients, with the following differences:
\begin{itemize}
  \item The orbital gradient $\mathbf{g^\Theta}$ in the CP-MCSCF equations (\ref{eqn:lagrange-mat}) or  (\ref{eqn:lagrange-mat-v}) is replaced by the generalized orbital gradient $\mathbf{T}(\Theta,\Lambda)$ with the average being calculated over two different states ($\Theta$ and $\Lambda$).
  \item In the expressions for the effective one- and two-particle RDMs (Eqs.~(\ref{eqn:oneptdm}) and (\ref{eqn:twoptdm})) the state-specific density matrices $\boldsymbol{\gamma^\Theta}$ and $\boldsymbol{\Gamma^\Theta}$ are replaced by the symmetrized transition density matrices $\boldsymbol{\gamma^{\Theta\Lambda}}$ and $\boldsymbol{\Gamma^{\Theta\Lambda}}$, respectively.
  \item For the calculation of the nonadiabatic couplings from the Lagrangian, Eq.~(\ref{eqn:sa-gradient}) is scaled by $(E_\Theta-E_\Lambda)^{-1}$ and an additional term enters, leading to 
  \begin{align}
   &\Braket{\Theta|\frac{\partial}{\partial x}| \Lambda} = \frac{1}{E_\Theta-E_\Lambda}\times \nonumber\\
   &\times\left(\sum_{pq}\frac{\partial\Braket{p|h|q}}{\partial x}\gamma^{\Theta,\text{e}}_{pq} + \frac{1}{2} \sum_{pqrs} \frac{\partial (pq|rs)}{\partial x}\Gamma^{\Theta,\text{e}}_{pqrs}- \right.\nonumber \\
   &- \left.\sum_{pq} X^{\Theta,\text{e}}_{pq} \frac{\partial S_{pq}}{\partial x}\right)+\sum_{pq}{}^\text{AS}f_{pq}, \label{eqn:nadc} \\
   \intertext{with}
    &^\text{AS}f_{pq} = - \frac{1}{2}\gamma^{\Theta\Lambda,\text{ns}}_{pq}\left((\frac{\partial \phi_p}{\partial x}|q) - (p|\frac{\partial \phi_q}{\partial x}) \right),
  \end{align}
  where $^\text{AS}f_{pq}$ is the contribution that arises from the antisymmetric transition density matrix $\gamma^{\Theta\Lambda,\text{ns}}_{pq}$  and the antisymmetric derivative overlap  $\left((\frac{\partial \phi_p}{\partial x}|q) - (p|\frac{\partial \phi_q}{\partial x}) \right)$\cite{Snyder_direct-compatible_JCP_2017}. In SA-CASSCF nonadiabatic coupling theory, this contribution is termed ``configuration state function (CSF) contribution''\cite{Fdez.Galvan_Analytical_JCTC_2016}, however here we prefer not to use this term as we do not operate in CSF basis in DMRG.
\end{itemize}

\section{Numerical examples}

\subsection{General computational details}

In the numerical examples below all SA-CASSCF calculations have been performed with the OpenMOLCAS\cite{OpenMOLCAS} software package, and the SA-DMRG-SCF calculations with the development version of the QCMaquis\cite{kell15a,kell16} DMRG program and the QCMaquis-OpenMOLCAS interface. In all examples, we have chosen the number of sweeps and the maximum number of renormalized block states $m$ in DMRG-SCF to approximate the CASSCF energy to an accuracy of $10^{-8}$\,Hartree. Although DMRG-SCF makes energy and gradient calculations for large active spaces accessible, in this work we restrict ourselves to a problem that can still be treated with the standard CASSCF for the sake of comparison.

For our first example, cyclobutadiene (Section~\ref{sec:cyclobutadiene}), a full-valence active space, consisting of 12 electrons in 12 orbitals, with Dunning's cc-pVDZ\cite{dunn89} basis set was chosen. The optimized orbitals in both SA-CASSCF and SA-DMRG-SCF calculations turned out to be identical, as expected, and the maximum number of renormalized block states $m$ in DMRG-SCF was set to 2000.

For our second example, the 1,2-dioxetanone (Sections~\ref{sec:dioxetanone}-\ref{sec:numerical}), an active space of 16 electrons in 13 orbitals, along with the ANO-RCC-VDZP basis set\cite{Roos_New_JPCA_2005} was employed as suggested in Ref.~\citenum{Liu_Theoretical_JACS_2009}. The maximum number of renormalized block states $m$ was raised to 5000, again to reproduce the CASSCF energy to an accuracy of $10^{-8}$\,Hartree.

\subsection{Cyclobutadiene: Dependence of the gradient on the choice of the linear response site}\label{sec:cyclobutadiene} In this subsection we examine how the choice of the linear response site determines the error in gradients and nonadiabatic couplings, with cyclobutadiene as an example. The automerization reaction of cyclobutadiene is the most prominent example of a process primarily driven by heavy-atom tunnelling\cite{Carpenter_Heavy-atom_JACS_1983,Carsky_Heavy-atom_TCA_1992,Schoonmaker_Quantum_JCP_2018}. As such, it has been subject to ab-initio dynamic simulations\cite{Eckert-Maksic_Automerization_JCP_2006}, which necessitate the usage of nuclear gradients. A single molecular structure along the automerization reaction path from Ref.~\citenum{Eckert-Maksic_Automerization_JCP_2006} has been chosen for our example below.

\onecolumngrid

\begin{figure}[t]
  \centering
  \includegraphics[width=0.85\textwidth]{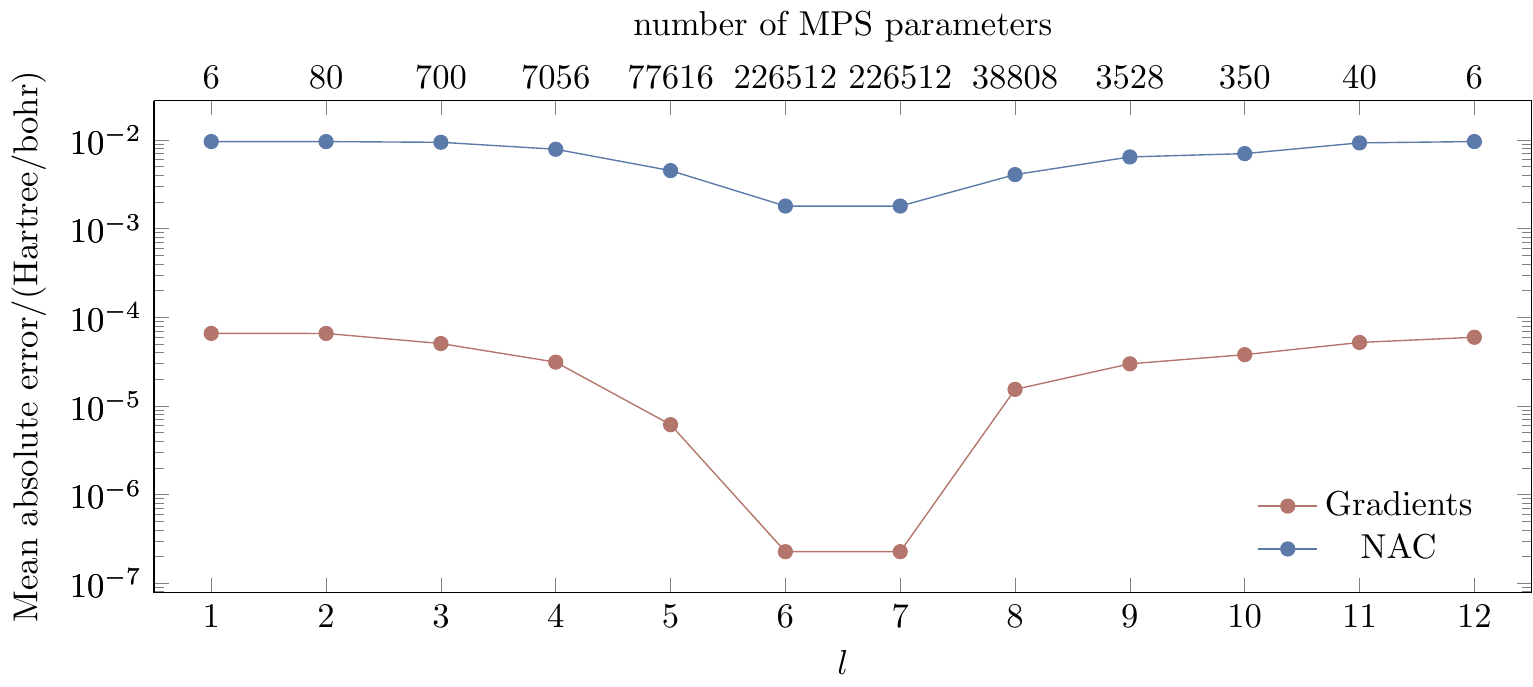}
  \caption{Mean absolute error (with respect to the SA-CASSCF result) of gradients for the S$_1$ state and the nonadiabatic couplings (NAC) between the S$_0$ and S$_1$ states of cyclobutadiene, calculated with SA-DMRG-SCF(12,12)[2000], for different linear response sites $l$. The $x$ axis on the top lists the corresponding number of MPS parameters: for $l=6$ and 7 the number of MPS parameters is equal to the number of the CI coefficients in the reference SA-CASSCF calculation.}\label{fig:lrsite}
\end{figure}
\ifpreprint{}{\twocolumngrid}

Figure~\ref{fig:lrsite} shows the mean absolute error of gradients of the S$_1$ state and the nonadiabatic couplings between the S$_0$ and S$_1$ states for various linear response sites, along with the corresponding number of MPS parameters. The mean absolute error is defined as
\begin{equation}
  \text{MAE} = \frac{1}{n}\sum_{i=1}^n \left| g_i^\text{DMRG-SCF} - g_i^\text{CASSCF}\right|\label{eq:mae}
\end{equation}
where $g_i^\text{DMRG-SCF}$ and $g_i^\text{CASSCF}$ are DMRG-SCF and CASSCF gradients or nonadiabatic couplings, respectively, and $n$ is their total number. Both errors clearly show their minimum in the middle of the lattice, where the number of MPS variational parameters of the wavefunction is the largest. Note that here, i.\,e., for sites 6 and 7, the number of MPS Lagrange multipliers is equal to the number of CI Lagrange multipliers in the CASSCF reference calculations, and therefore one would expect that in case of a perfect match of CASSCF and DMRG-SCF wavefunctions the gradients should also be identical. Still, we observe a small error of approximately $10^{-7}$\,Hartree/bohr, which is attributed to the small numerical differences between CASSCF and DMRG-SCF wavefunctions, caused by nonzero convergence thresholds for the orbital optimization, DMRG sweeps, and the PCG method in CP-DMRG-SCF equations. By tightening the convergence thresholds the error decreases. Consequently, for the subsequent examples, we chose a linear response site in the middle of the lattice for the best accuracy.

With linear response sites towards the edge of the lattice, the gradient error increases by up to three orders of magnitude. This error arises from our single-site MPS parametrization and the BVOE because, as we already remarked above, tensors at the edges of the lattice have fewer elements leading to fewer MPS parameters. However, by varying the linear response site we may trade the gradient accuracy against the computational cost, as CP-DMRG-SCF equations with fewer Lagrange multipliers converge faster. In the middle of the lattice in our example the number of MPS parameters is equal to the number of CSFs in a CASSCF calculation, hence spanning the same variational space as in the CASSCF wavefunction: the BVOE approaches zero in this case. For SA-DMRG-SCF wavefunctions that approximate SA-CASSCF, as we will see in Section~\ref{sec:numerical}, BVOE is nonzero even for linear response sites in the middle of the lattice, but the error is still small enough for practical purposes.

Avoiding BVOE completely to minimize the DMRG-SCF gradient error would require a parametrization that comprises MPS parameters from all sites: either a tangent-space based approach\cite{Wouters_Thouless_PRB_2013,Nakatani_Linear_JCP_2014} or an extension of our current scheme to a sweep-like procedure, where CP-DMRG-SCF equations are solved for every site and the nonvariational contributions to the RDMs are obtained at the end of the sweep. Both procedures are subject of our future work, but we would expect both of them to have a very high computational cost either due to the sweep procedure or employing a very large number of MPS parameters. Yet, the current algorithm based on a single linear response site, as we see here, already achieves a very good accuracy at a rather moderate computational cost by solving the CP-DMRG-SCF equations only once.

Surprisingly, the nonadiabatic couplings show errors up to four orders of magnitude larger than the gradients. However, the error in nonadiabatic couplings can be attributed to their larger sensitivity to the wavefunction quality in general: in a DMRG-SCF calculation with an extremely tight energy convergence threshold for the sweep of $10^{-12}$\,Hartree the average deviation of the couplings from their SA-CASSCF counterparts could be reduced to $7\times10^{-6}$\,Hartree/bohr. The sensitivity of nonadiabatic couplings to the quality of the reference wavefunction is not limited to DMRG-SCF but can be observed between several reference CASSCF calculations: a SA-CASSCF calculation with a lower Davidson diagonalization threshold of $10^{-9}$\,Hartree yields nonadiabatic couplings which differ on average by $1.3\times10^{-7}$\,Hartree/bohr from the original CASSCF calculation.
Note that these deviations in the nonadiabatic couplings have no practical effect on a conical intersection optimization, as we will see below.

\subsection{1,2-Dioxetanone: Optimization of a conical intersection}\label{sec:dioxetanone}

1,2-dioxetanone (Fig.~\ref{fig:diox}a) is a simple chemoluminescent compound, whose thermal dissociation mechanism, along with those of its substituted derivatives have been extensively studied in chemo- and bioluminescence studies.\cite{Schmidt_Kinetics_JACS_1978,Adam_Cyclic_JACS_1979,Turro_Chemiluminescent_JACS_1980,Liu_Theoretical_JACS_2009,Liu_CASSCFCASPT2_CPL_2009,Navizet_Chemistry_C_2011,Yue_Theoretical_JPCL_2015,Farahani_combined_RA_2017,Vacher_Chemi-_CR_2018,Yue_Two_JCTC_2019} \citet{Liu_Theoretical_JACS_2009} performed an extensive computational study on the dissociation pathway of 1,2-dioxetanone and located two conical intersections between the $\sigma\sigma^*$ and the $n\sigma^*$ states along the pathway. In the vicinity of these conical intersections both states share a significant biradical character, which mandates a multiconfigurational treatment such as CASSCF. We performed an optimization of the first ($\sigma\sigma^*$),($n\sigma^*$) conical intersections along the reaction path, named by \citet{Liu_Theoretical_JACS_2009} as ``$^1\sigma\sigma^*$-TS'', both with CASSCF and DMRG-SCF. The optimization run for both methods is presented in Table~\ref{tab:ciopt}. At each step the energies and the gradients are nearly identical, and the discrepancies are very small -- below $2.8\times10^{-6}$\,Hartree for the energies, $6\times10^{-5}$\,Hartree/bohr for gradient norms and $6.3\times10^{-5}$\,Hartree/bohr for the maximum gradient element -- in all cases well below the convergence thresholds for the optimization. These small discrepancies again decrease even further by tightening the convergence thresholds. The optimized structures are also essentially identical for both methods: the maximum absolute difference in the Cartesian coordinates is $2\times10^{-4}$\,\AA{} for both optimized structures. The most important bond lengths and angles of the optimized structure are presented in Fig.~\ref{fig:diox}b.
As we see, a conical intersection optimization with our SA-DMRG-SCF analytical gradient and nonadiabatic coupling \emph{ansatz} is able to accurately reproduce the SA-CASSCF optimization result.

\onecolumngrid

  \begin{table}[ht]
    \caption{Average electronic energy, gradient norm, and the maximum gradient element throughout the ($\sigma\sigma^*$),($n\sigma^*$) conical intersection optimization in 1,2-dioxetanone.}\label{tab:ciopt}
    \begin{tabular}{ccc|cc|rr}
      \toprule
        Step & \multicolumn{2}{c}{Electronic energy/} & \multicolumn{2}{c}{Avg.\ gradient\ norm/} & \multicolumn{2}{c}{Max.\ gradient element/} \\
            & \multicolumn{2}{c}{Hartree} & \multicolumn{2}{c}{Hartree/bohr}& \multicolumn{2}{c}{Hartree/bohr}\\
        \colrule
        & CASSCF & DMRG-SCF & CASSCF & DMRG-SCF & CASSCF & DMRG-SCF \\
        \cline{2-3} \cline{4-5} \cline{6-7}
    1  & $-301.76392482$ & $-301.76392481$ & $0.098030$ & $0.098090$ & $-0.070811$ & $-0.070874$ \\
    2  & $-301.77670336$ & $-301.77670325$ & $0.089606$ & $0.089626$ & $-0.050384$ & $-0.050435$ \\
    3  & $-301.79392062$ & $-301.79391779$ & $0.159354$ & $0.159318$ & $ 0.127371$ & $ 0.127362$ \\
    4  & $-301.78862423$ & $-301.78862694$ & $0.073202$ & $0.073216$ & $-0.034099$ & $-0.034124$ \\
    5  & $-301.81089953$ & $-301.81089903$ & $0.066443$ & $0.066451$ & $ 0.037325$ & $ 0.037307$ \\
    6  & $-301.81982062$ & $-301.81982236$ & $0.027812$ & $0.027800$ & $-0.017275$ & $-0.017280$ \\
    7  & $-301.82254958$ & $-301.82254963$ & $0.017071$ & $0.017075$ & $ 0.012606$ & $ 0.012609$ \\
    8  & $-301.82398804$ & $-301.82398842$ & $0.011692$ & $0.011693$ & $-0.008136$ & $-0.008114$ \\
    9  & $-301.82423076$ & $-301.82423099$ & $0.004222$ & $0.004204$ & $-0.003256$ & $-0.003232$ \\
   10  & $-301.82426240$ & $-301.82426246$ & $0.000897$ & $0.000897$ & $-0.000552$ & $-0.000549$ \\
   11  & $-301.82426314$ & $-301.82426325$ & $0.000404$ & $0.000387$ & $-0.000269$ & $-0.000254$ \\
        \botrule
    \end{tabular}
  \end{table}
  \ifpreprint{}{\twocolumngrid}

\subsection{1,2-Dioxetanone: Gradient convergence with number of renormalized block states}\label{sec:dioxetanone-conv}
\begin{figure}[t]
  \centering
  \setlength{\tabcolsep}{1em}
  \begin{tabular}{cc}
    \includegraphics[width=0.1\textwidth]{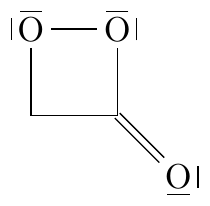} & \includegraphics[width=0.15\textwidth]{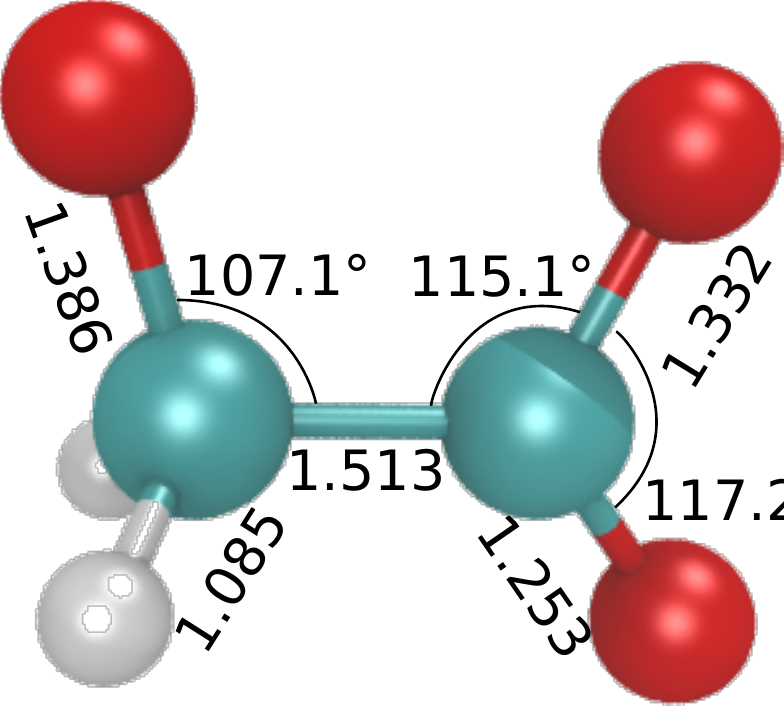} \\
    (a) & (b)
  \end{tabular}
  \caption{(a): Lewis structure of 1,2-dioxetanone. (b): ($\sigma\sigma^*$),($n\sigma^*$)-CI (named ``$^1\sigma\sigma^*$-TS'' in Ref.~\citenum{Liu_Theoretical_JACS_2009}) structure optimized in this work with most important bond lengths (in \AA) and angles. }\label{fig:diox}
\end{figure}

\begin{figure}[h]
  \centering
  \includegraphics[width=0.48\textwidth]{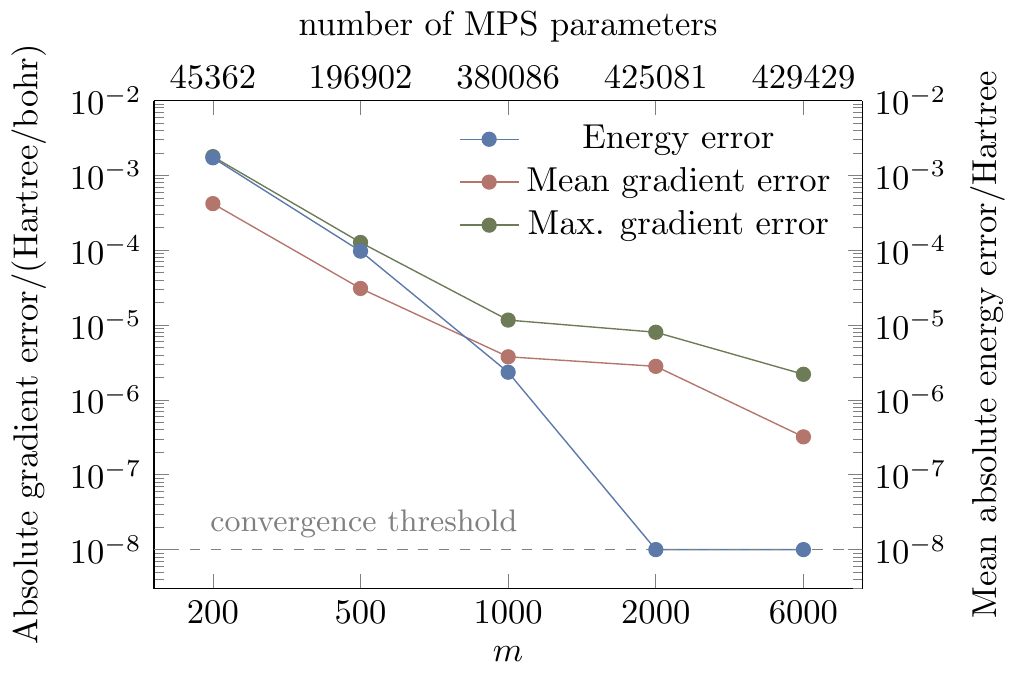}
  \caption{Mean and maximum absolute errors (with respect to the SA-CASSCF result) of gradients for the S$_1$ state for 1,2-dioxetanone at a structure in the vicinity of ($\sigma\sigma^*$),($n\sigma^*$) conical intersection, from SA-DMRG-SCF(16,13) calculations with different values for the number of renormalized block states $m$. For $m=6000$, the number of MPS parameters corresponds to that of the reference SA-CASSCF calculation. The average error for the energies of S$_0$ and S$_1$ states is shown for comparison.}\label{fig:m}
\end{figure}
  Having established that our SA-DMRG-SCF gradient \emph{ansatz} is able to reproduce SA-CASSCF gradients to an accuracy that is only dependent on convergence thresholds in cases where an $m$ value is sufficiently large to reproduce the SA-CASSCF wavefunction to arbitrary accuracy, we now study the gradient error for smaller $m$ values, i.\,e., when the approximation of the SA-CASSCF wavefunction by a SA-DMRG-SCF wavefunction is of a reduced quality. For this, a gradient calculation has been performed for a structure in the vicinity of the ($\sigma\sigma^*$),($n\sigma^*$) conical intersection with SA-DMRG-SCF(16,13), but with several $m$ values varying from 200 to 6000.

  Fig.~\ref{fig:m} shows the mean and maximum absolute gradient error for each $m$, along with the average energy error. We recognize that for large $m$ values ($m \ge 2000$), the gradient error is one to two orders of magnitude larger than the energy error, and that an $m$ value which is sufficient to converge the energy up to an error of $10^{-8}$\,Hartree is not sufficient to converge the gradient to the same accuracy as the energy. Even a much larger value ($m=6000$) does not converge the gradient to the same accuracy, although the error decreases by another order of magnitude. We expect that tightening the convergence threshold for the energy along with larger $m$ values would improve the convergence even further. However (as we see both from this result and the conical intersection optimization in the previous section) even at $m=2000$, the gradient error is well below typical convergence thresholds for optimizations and can therefore be neglected for practical applications.

  However, for small $m$ values we see an interesting development: although the maximum gradient error is approximately the same or larger than the energy error, the mean gradient error is actually smaller. The crossover of the mean energy and gradient errors occurs around $m=1000$. While we believe that this result is caused by a fortituous error cancellation, it implies that for typical DMRG-SCF calculations, which are performed for active spaces inaccessible by standard CASSCF the accuracy of the gradient error is of the same order of magnitude or smaller than the energy error. In other words, in these calculations we may extract energies and gradients of a similar accuracy from the same wavefunction without further refinements such as imposing tighter convergence thresholds.

\subsection{1,2-Dioxetanone: Comparison with numerical gradients}\label{sec:numerical}

\begin{table}[h]
  \caption{Mean absolute error of gradient elements (in Hartree/bohr) calculated with SA-DMRG-SCF ($m=500$) and SA-CASSCF, both analytically and numerically. The first two entries are MAE as defined in Eq.~(\ref{eq:mae}) for analytical and numerical gradients, respectively, while the last two entries are the deviation between analytical and numerical SA-DMRG-SCF and SA-CASSCF gradients, respectively.}\label{tab:numeric}
  \begin{tabular}{cc}
    \toprule
             & DMRG-SCF$-$CASSCF \\
    \colrule
    Analytic & $1.9\times10^{-3}$       \\
    Numerical & $2.2\times10^{-4}$       \\
    \colrule
             & Analytic$-$Numerical \\
    \colrule
    DMRG-SCF & $2.0\times10^{-3}$ \\
    CASSCF & $5.6\times10^{-5}$\\
    \botrule
  \end{tabular}
\end{table}

To further assess the errors of the SA-DMRG-SCF analytical gradient for small $m$ values, in particular to quantify the BVOE, we compared the accuracy of the analytical gradient to a numerical gradient obtained for the same $m$ value. For this comparison, again a (16,13) active space and an $m$ value of 500 was chosen. Although it is beyond the scope of this work to provide many examples, we attempted to employ several testcases, and, as such, here we calculated the S$_1$ excited state gradient for the S$_0$ optimized structure (rather than a structure in the vicinity of the conical intersection as in Section~\ref{sec:dioxetanone-conv}).
For the numerical gradient, a convergence with the step size has been verified by repeating the numerical gradient calculations with step sizes of 0.0025 and 0.00125\,bohr and ensuring that the difference in every gradient component does not exceed $10^{-7}$\,Hartree/bohr. Since even a numerical gradient that is fully converged with respect to the step size may deviate from the analytical gradient, a comparison between the CASSCF analytical and numerical gradient was considered to quantify this error. Table~\ref{tab:numeric} lists the mean absolute deviations between numerical and analytical gradients, both for DMRG-SCF and CASSCF. From the comparison of the differences between analytical and numerical  gradients we estimate the BVOE to be about $1.9\times{}10^{-3}$\,Hartree/bohr (see first entry in Table~\ref{tab:numeric}). It is an order of magnitude larger than the gradient error which arises due to the small $m$ value alone (equal to the mean absolute deviation between numerical DMRG-SCF and CASSCF gradients, $2.2\times{}10^{-4}$\,Hartree/bohr).

It is noteworthy that the BVOE in this example is approximately one order of magnitude larger than the analogous error for $m=500$ in Fig.~\ref{fig:m}. This is not surprising as one would expect the error to decrease when the non-variational MPS contribution to the gradient decreases. In the vicinity of a conical intersection $S_0$ and $S_1$ wavefunctions may mix and thus their electronic character becomes more similar. Under these circumstances one would expect a smaller non-variational contribution to the gradient. However, at the S$_0$ minimum the S$_0$ and S$_1$ wavefunctions are very different, which in turn would lead to a larger non-variational contribution, and, consequently, a larger BVOE. While this shows that our approximation is particularly well suited for the optimization of conical intersections, it is the scope of future work to assess our scheme for other photochemical applications, such as excited state structure optimizations or surface hopping dynamics.

\section{Conclusions}\label{sec:conclusions}

In this paper, we presented an implementation of a Lagrangian-based \emph{ansatz} for the analytical state-average DMRG-SCF gradients and nonadiabatic couplings which requires construction and solution of CP-MCSCF equations. Our \emph{ansatz} generalizes the SA-CASSCF gradients theory first presented by \citet{Jones2001Analytical} in 2001 to wavefunctions encoded as MPSs. We derive the Lagrange multipliers in the CP-MCSCF equations from the mixed-canonical representation of the MPS wavefunction at one particular site, here called the linear response site, which can be chosen arbitrarily. The choice of the linear response site can be made a tradeoff between accuracy and computational cost. As it is, such an MPS parametrization constitutes an approximation, because we consider the variation of the MPS at one particular site only. However, as the MPS approaches the full CI wavefunction, the error associated with this approximation approaches zero. We further argue that for the validity of our definition of MPS Lagrange multipliers the MPS for all states must be optimized simultaneously with common left and right boundaries.
Finally, we showed that our SA-DMRG-SCF gradient and nonadiabatic coupling \emph{ansatz} will exactly reproduce SA-CASSCF gradients and nonadiabatic couplings, respectively, if the SA-DMRG-SCF wavefunction reproduces the SA-CASSCF wavefunction exactly. If this is not the case, both gradients and, in particular, nonadiabatic couplings show errors that are larger than the errors in the energies. This is largely due to the approximate nature of our MPS parametrization, however, this error is still small enough for practical purposes, especially for conical intersection optimizations. We have demonstrated the feasibility of our method in practical applications by performing a conical intersection optimization of 1,2-dioxetanone, reproducing the result of a SA-CASSCF conical intersection optimization. Our development, in addition to applications for conical intersection optimization for systems with large active spaces, paves the way for surface hopping studies and other excited state studies with DMRG.

\section*{Acknowledgments}

L.F.\ acknowledges the Austrian Science Fund for a Schr\"{o}dinger fellowship (Project No.\ J 3935).
This work was supported by the Schweizerischer Nationalfonds (SNF project 200021\_182400), National Natural Science Foundation of China (NO. 21703260), ETH Z\"urich (ETH Fellowship No. FEL-49 18-1) and the Informatization Program of the Chinese Academy of Sciences (NO. XXH13506-403). We thank Prof.\ Roland Lindh and Dr.\ Christopher Stein for fruitful discussions.

\section*{References}
%\bibliographystyle{achemso2}
%\bibliography{dmrgscf-gradient}

\begin{thebibliography}{100}

\bibitem{Pierloot_Calculations_TaCC_2005}
Pierloot,~K.  Calculations of {{Electronic Spectra}} of {{Transition Metal
  Complexes}}.   In  \textit{Theoretical and {{Computational Chemistry}}},
  Vol.~16; Olivucci,~M.,\ \ Ed.;  {Elsevier}: 2005.

\bibitem{Neese_Advanced_CCR_2007}
Neese,~F.;\ \ Petrenko,~T.;\ \ Ganyushin,~D.;\ \ Olbrich,~G.  Advanced Aspects
  of Ab Initio Theoretical Optical Spectroscopy of Transition Metal Complexes:
  {{Multiplets}}, Spin-Orbit Coupling and Resonance {{Raman}} Intensities,
  \textit{Coord. Chem. Rev.} \textbf{2007,} \textsl{251,} 288-327.

\bibitem{Gagliardi_Transition_RiCC_2007}
Gagliardi,~L.  Transition {{Metal}}- and {{Actinide}}-{{Containing Systems
  Studied}} with {{Multiconfigurational Quantum Chemical Methods}}.   In
  \textit{Reviews in {{Computational Chemistry}}}; Lipkowitz,~K.~B.;\ \
  Cundari,~T.~R.,\ \ Eds.;  {John Wiley \& Sons, Inc.}: 2007.

\bibitem{Neese_Correlated_JBIC_2011}
Neese,~F.;\ \ Liakos,~D.~G.;\ \ Ye,~S.  Correlated Wavefunction Methods in
  Bioinorganic Chemistry,  \textit{J. Biol. Inorg. Chem.} \textbf{2011,}
  \textsl{16,} 821-829.

\bibitem{Pierloot_Transition_IJQC_2011}
Pierloot,~K.  Transition Metals Compounds: {{Outstanding}} Challenges for
  Multiconfigurational Methods,  \textit{Int. J. Quantum Chem.} \textbf{2011,}
  \textsl{111,} 3291-3301.

\bibitem{Daniel_Photochemistry_CCR_2015}
Daniel,~C.  Photochemistry and Photophysics of Transition Metal Complexes:
  {{Quantum}} Chemistry,  \textit{Coord. Chem. Rev.} \textbf{2015,}
  \textsl{282\textendash{}283,} 19-32.

\bibitem{Ashley_Ironing_CCR_2017}
Ashley,~D.~C.;\ \ Jakubikova,~E.  Ironing out the Photochemical and
  Spin-Crossover Behavior of {{Fe}}({{II}}) Coordination Compounds with
  Computational Chemistry,  \textit{Coord. Chem. Rev.} \textbf{2017,}
  \textsl{337,} 97-111.

\bibitem{Radon_Benchmarking_PCCP_2019}
Rado\'n,~M.  Benchmarking Quantum Chemistry Methods for Spin-State Energetics
  of Iron Complexes against Quantitative Experimental Data,  \textit{Phys.
  Chem. Chem. Phys.} \textbf{2019,} \textsl{21,} 4854-4870.

\bibitem{Vogiatzis_Computational_CR_2019}
Vogiatzis,~K.~D.;\ \ Polynski,~M.~V.;\ \ Kirkland,~J.~K.;\ \ Townsend,~J.;\ \
  Hashemi,~A.;\ \ Liu,~C.;\ \ Pidko,~E.~A.  Computational {{Approach}} to
  {{Molecular Catalysis}} by 3d {{Transition Metals}}: {{Challenges}} and
  {{Opportunities}},  \textit{Chem. Rev.} \textbf{2019,} \textsl{119,}
  2453-2523.

\bibitem{Gonzalez_Progress_C_2012}
Gonz\'alez,~L.;\ \ Escudero,~D.;\ \ Serrano-Andr\'es,~L.  Progress and
  {{Challenges}} in the {{Calculation}} of {{Electronic Excited States}},
  \textit{ChemPhysChem} \textbf{2012,} \textsl{13,} 28--51.

\bibitem{Plasser_Electronically_TCA_2012}
Plasser,~F.;\ \ Barbatti,~M.;\ \ Aquino,~A. J.~A.;\ \ Lischka,~H.
  Electronically Excited States and Photodynamics: A Continuing Challenge,
  \textit{Theor. Chem. Acc.} \textbf{2012,} \textsl{131,} 1-14.

\bibitem{Ghosh_Combining_CR_2018}
Ghosh,~S.;\ \ Verma,~P.;\ \ Cramer,~C.~J.;\ \ Gagliardi,~L.;\ \ Truhlar,~D.~G.
  Combining {{Wave Function Methods}} with {{Density Functional Theory}} for
  {{Excited States}},  \textit{Chem. Rev.} \textbf{2018,} \textsl{118,}
  7249-7292.

\bibitem{Lischka_Multireference_CR_2018}
Lischka,~H.;\ \ Nachtigallov\'a,~D.;\ \ Aquino,~A. J.~A.;\ \ Szalay,~P.~G.;\ \
  Plasser,~F.;\ \ Machado,~F. B.~C.;\ \ Barbatti,~M.  Multireference
  {{Approaches}} for {{Excited States}} of {{Molecules}},  \textit{Chem. Rev.}
  \textbf{2018,} \textsl{118,} 7293-7361.

\bibitem{Serrano-Andres_Are_JPPCPR_2009}
Serrano-Andr\'es,~L.;\ \ Merch\'an,~M.  Are the Five Natural {{DNA}}/{{RNA}}
  Base Monomers a Good Choice from Natural Selection?: {{A}} Photochemical
  Perspective,   \textbf{2009,} \textsl{10,} 21-32.

\bibitem{Middleton_DNA_ARPC_2009}
Middleton,~C.~T.;\ \ de~La~Harpe,~K.;\ \ Su,~C.;\ \ Law,~Y.~K.;\ \
  Crespo-Hern\'andez,~C.~E.;\ \ Kohler,~B.  {{DNA Excited}}-{{State Dynamics}}:
  {{From Single Bases}} to the {{Double Helix}},  \textit{Annu. Rev. Phys.
  Chem.} \textbf{2009,} \textsl{60,} 217-239.

\bibitem{Mai_Excitation_2014}
Mai,~S.;\ \ Richter,~M.;\ \ Marquetand,~P.;\ \ Gonz\'alez,~L.  Excitation of
  {{Nucleobases}} from a {{Computational Perspective II}}: {{Dynamics}},
  \textit{Top. Curr. Chem.} \textbf{2015,}  99-154.

\bibitem{Gust_Solar_ACR_2009}
Gust,~D.;\ \ Moore,~T.~A.;\ \ Moore,~A.~L.  Solar {{Fuels}} via {{Artificial
  Photosynthesis}},  \textit{Acc. Chem. Res.} \textbf{2009,} \textsl{42,}
  1890-1898.

\bibitem{Gust_Realizing_FD_2012}
Gust,~D.;\ \ Moore,~T.~A.;\ \ Moore,~A.~L.  Realizing Artificial
  Photosynthesis,  \textit{Faraday Discuss.} \textbf{2012,} \textsl{155,} 9-26.

\bibitem{Jager_Using_CCR_2015}
J\"ager,~M.;\ \ Freitag,~L.;\ \ Gonz\'alez,~L.  Using Computational Chemistry
  to Design {{Ru}} Photosensitizers with Directional Charge Transfer,
  \textit{Coord. Chem. Rev.} \textbf{2015,} \textsl{304\textendash{}305,}
  146-165.

\bibitem{Agbe_Recent_JIEC_2019}
Agbe,~H.;\ \ Nyankson,~E.;\ \ Raza,~N.;\ \ Dodoo-Arhin,~D.;\ \ Chauhan,~A.;\ \
  Osei,~G.;\ \ Kumar,~V.;\ \ Kim,~K.-H.  Recent Advances in Photoinduced
  Catalysis for Water Splitting and Environmental Applications,  \textit{J.
  Ind. Eng. Chem.} \textbf{2019,} \textsl{72,} 31-49.

\bibitem{Dolmans_Photodynamic_NRC_2003}
Dolmans,~D. E. J. G.~J.;\ \ Fukumura,~D.;\ \ Jain,~R.~K.  Photodynamic Therapy
  for Cancer,  \textit{Nat. Rev. Cancer} \textbf{2003,} \textsl{3,} 380-387.

\bibitem{Rose_Fiat_COCB_2008}
Rose,~M.~J.;\ \ Mascharak,~P.~K.  Fiat {{Lux}}: Selective Delivery of High Flux
  of Nitric Oxide ({{NO}}) to Biological Targets Using Photoactive Metal
  Nitrosyls,  \textit{Curr. Opin. Chem. Biol.} \textbf{2008,} \textsl{12,}
  238-244.

\bibitem{Ormond_Dye_M_2013}
Ormond,~A.~B.;\ \ Freeman,~H.~S.  Dye {{Sensitizers}} for {{Photodynamic
  Therapy}},  \textit{Materials} \textbf{2013,} \textsl{6,} 817-840.

\bibitem{Mehraban_Developments_M_2015}
Mehraban,~N.;\ \ Freeman,~H.~S.  Developments in {{PDT Sensitizers}} for
  {{Increased Selectivity}} and {{Singlet Oxygen Production}},
  \textit{Materials} \textbf{2015,} \textsl{8,} 4421-4456.

\bibitem{roos80}
Roos,~B.~O.;\ \ Taylor,~P.~R.;\ \ Siegbahn,~P.~E.  {A complete active space
  (SCF) method (CASSCF) using a density matrix formulated super-CI approach},
  \textit{Chem. Phys.} \textbf{1980,} \textsl{48,} 157-173.

\bibitem{sieg81}
Siegbahn,~P.~E.;\ \ Alml{\"o}f,~J.;\ \ Heiberg,~A.;\ \ Roos,~B.~O.  {The
  complete active space SCF (CASSCF) method in a Newton--Raphson formulation
  with application to the HNO molecule},  \textit{J. Chem. Phys.}
  \textbf{1981,} \textsl{74,} 2384-2396.

\bibitem{shep87}
Shepard,~R.  {The multiconfiguration self-consistent field method},
  \textit{Adv. Chem. Phys.} \textbf{1987,} \textsl{67,} 63-200.

\bibitem{olse11}
Olsen,~J.  The {CASSCF} method: A perspective and commentary,  \textit{Int. J.
  Quantum Chem.} \textbf{2011,} \textsl{111,} 3267--3272.

\bibitem{aqui15a}
Aquilante,~F. \textit{et al.}\   Molcas 8: New capabilities for
  multiconfigurational quantum chemical calculations across the periodic table,
   \textit{J. Comput. Chem.} \textbf{2016,} \textsl{37,} 506-541.

\bibitem{Vogiatzis_Pushing_JCP_2017}
Vogiatzis,~K.~D.;\ \ Ma,~D.;\ \ Olsen,~J.;\ \ Gagliardi,~L.;\ \ de~Jong,~W.~A.
  Pushing Configuration-Interaction to the Limit: {{Towards}} Massively
  Parallel {{MCSCF}} Calculations,  \textit{J. Chem. Phys.} \textbf{2017,}
  \textsl{147,} 184111.

\bibitem{White1998Ab}
White,~S.~R.;\ \ Martin,~R.~L.  Ab initio quantum chemistry using the density
  matrix renormalization group,  \textit{J. Chem. Phys.} \textbf{1998,}
  \textsl{110,} 4127-4130.

\bibitem{mitrushenkov2001quantum}
Mitrushenkov,~A.~O.;\ \ Fano,~G.;\ \ Ortolani,~F.;\ \ Linguerri,~R.;\ \
  Palmieri,~P.  Quantum chemistry using the density matrix renormalization
  group,  \textit{J. Chem. Phys.} \textbf{2001,} \textsl{115,} 6815-6821.

\bibitem{chan2002highly}
Chan,~G. K.-L.;\ \ Head-Gordon,~M.  Highly correlated calculations with a
  polynomial cost algorithm: A study of the density matrix renormalization
  group,  \textit{J. Chem. Phys.} \textbf{2002,} \textsl{116,} 4462-4476.

\bibitem{legeza2003controlling}
Legeza,~{\"O}.;\ \ R{\"o}der,~J.;\ \ Hess,~B.  Controlling the accuracy of the
  density-matrix renormalization-group method: The dynamical block state
  selection approach,  \textit{Phys. Rev. B} \textbf{2003,} \textsl{67,}
  125114.

\bibitem{chan2004algorithm}
Chan,~G. K.-L.  An algorithm for large scale density matrix renormalization
  group calculations,  \textit{J. Chem. Phys.} \textbf{2004,} \textsl{120,}
  3172-3178.

\bibitem{moritz2005convergence}
Moritz,~G.;\ \ Hess,~B.~A.;\ \ Reiher,~M.  Convergence behavior of the
  density-matrix renormalization group algorithm for optimized orbital
  orderings,  \textit{J. Chem. Phys.} \textbf{2005,} \textsl{122,} 024107.

\bibitem{moritz2005relativistic}
Moritz,~G.;\ \ Wolf,~A.;\ \ Reiher,~M.  Relativistic DMRG calculations on the
  curve crossing of cesium hydride,  \textit{J. Chem. Phys.} \textbf{2005,}
  \textsl{123,} 184105.

\bibitem{rissler2006measuring}
Rissler,~J.;\ \ Noack,~R.~M.;\ \ White,~S.~R.  Measuring orbital interaction
  using quantum information theory,  \textit{Chem. Phys.} \textbf{2006,}
  \textsl{323,} 519-531.

\bibitem{mcculloch2007density}
McCulloch,~I.~P.  From density-matrix renormalization group to matrix product
  states,  \textit{J. Stat. Mech-Theory E.} \textbf{2007,} \textsl{2007,}
  P10014.

\bibitem{lege08}
Legeza,~{\"O}.;\ \ Noack,~R.~M.;\ \ Solyom,~J.;\ \ Tincani,~L.  Applications of
  quantum information in the density-matrix renormalization group.   In
  \textit{Computational Many-Particle Physics}, Vol.~739; Fehske,~H.;\ \
  Schneider,~R.;\ \ Weisse,~A.,\ \ Eds.;  2008.

\bibitem{chan2009density}
Chan,~G. K.-L.;\ \ Zgid,~D.  The density matrix renormalization group in
  quantum chemistry,  \textit{Ann. Rep. Comput. Chem.} \textbf{2009,}
  \textsl{5,} 149--162.

\bibitem{mart10}
Marti,~K.~H.;\ \ Reiher,~M.  {The Density Matrix Renormalization Group
  Algorithm in Quantum Chemistry},  \textit{Z. Phys. Chem.} \textbf{2010,}
  \textsl{224,} 583-599.

\bibitem{schollwock2011density}
Schollw{\"o}ck,~U.  The density-matrix renormalization group in the age of
  matrix product states,  \textit{Ann. Phys.} \textbf{2011,} \textsl{326,}
  96-192.

\bibitem{chan2011density}
Chan,~G. K.-L.;\ \ Sharma,~S.  The density matrix renormalization group in
  quantum chemistry,  \textit{Ann. Rev. Phys. Chem.} \textbf{2011,}
  \textsl{62,} 465-481.

\bibitem{Kurashige2009High}
Kurashige,~Y.;\ \ Yanai,~T.  High-performance ab initio density matrix
  renormalization group method: applicability to large-scale multireference
  problems for metal compounds.,  \textit{J. Chem. Phys.} \textbf{2009,}
  \textsl{130,} 234114.

\bibitem{Knecht2014Communication}
Knecht,~S.;\ \ Örs Legeza,;\ \ Reiher,~M.  Communication: Four-component
  density matrix renormalization group,  \textit{J. Chem. Phys.} \textbf{2014,}
  \textsl{140,} 041101.

\bibitem{wout14}
Wouters,~S.;\ \ van Neck,~D.  The density matrix renormalization group for ab
  initio quantum chemistry,  \textit{Eur. Phys. J. D} \textbf{2014,}
  \textsl{68,} 272.

\bibitem{Yanai2015Density}
Yanai,~T.;\ \ Kurashige,~Y.;\ \ Mizukami,~W.;\ \ Chalupský,~J.;\ \
  Lan,~T.~N.;\ \ Saitow,~M.  Density matrix renormalization group for ab initio
  Calculations and associated dynamic correlation methods: A review of theory
  and applications,  \textit{Int. J. Quantum Chem.} \textbf{2015,}
  \textsl{115,} 283-299.

\bibitem{knec16a}
Knecht,~S.;\ \ Hedeg\aa{}rd,~E.~D.;\ \ Keller,~S.;\ \ Kovyrshin,~A.;\ \
  Ma,~Y.;\ \ Muolo,~A.;\ \ Stein,~C.~J.;\ \ Reiher,~M.  New approaches for ab
  initio calculations of molecules with strong electron correlation,
  \textit{Chimia} \textbf{2016,} \textsl{70,} 244--251.

\bibitem{Chan2016Matrix}
Chan,~G.~K.;\ \ Keselman,~A.;\ \ Nakatani,~N.;\ \ Li,~Z.;\ \ White,~S.~R.
  Matrix product operators, matrix product states, and ab initio density matrix
  renormalization group algorithms,  \textit{J. Chem. Phys.} \textbf{2016,}
  \textsl{145,} 2863-2865.

\bibitem{white1992density}
White,~S.~R.  Density matrix formulation for quantum renormalization groups,
  \textit{Phys. Rev. Lett.} \textbf{1992,} \textsl{69,} 2863.

\bibitem{white1992real}
White,~S.~R.;\ \ Noack,~R.  Real-space quantum renormalization groups,
  \textit{Phys. Rev. Lett.} \textbf{1992,} \textsl{68,} 3487.

\bibitem{White_initio_JCP_1999}
White,~S.~R.;\ \ Martin,~R.~L.  Ab Initio Quantum Chemistry Using the Density
  Matrix Renormalization Group,  \textit{J. Chem. Phys.} \textbf{1999,}
  \textsl{110,} 4127-4130.

\bibitem{Daul_Full-CI_IJQC_2000}
Daul,~S.;\ \ Ciofini,~I.;\ \ Daul,~C.;\ \ White,~S.~R.  Full-{{CI}} Quantum
  Chemistry Using the Density Matrix Renormalization Group,  \textit{Int. J.
  Quantum Chem.} \textbf{2000,} \textsl{79,} 331-342.

\bibitem{Mitrushenkov_Quantum_JCP_2001}
Mitrushenkov,~A.~O.;\ \ Fano,~G.;\ \ Ortolani,~F.;\ \ Linguerri,~R.;\ \
  Palmieri,~P.  Quantum Chemistry Using the Density Matrix Renormalization
  Group,  \textit{J. Chem. Phys.} \textbf{2001,} \textsl{115,} 6815-6821.

\bibitem{Mitrushenkov_Quantum_JCP_2003}
Mitrushenkov,~A.~O.;\ \ Linguerri,~R.;\ \ Palmieri,~P.;\ \ Fano,~G.  Quantum
  Chemistry Using the Density Matrix Renormalization Group {{II}},  \textit{J.
  Chem. Phys.} \textbf{2003,} \textsl{119,} 4148-4158.

\bibitem{Mitrushchenkov_importance_IJQC_2011}
Mitrushchenkov,~A.~O.;\ \ Fano,~G.;\ \ Linguerri,~R.;\ \ Palmieri,~P.  On the
  Importance of Orbital Localization in {{QC}}-{{DMRG}} Calculations,
  \textit{Int. J. Quantum Chem.} \textbf{2011,} \textsl{112,} 1606-1619.

\bibitem{Chan_Highly_JCP_2002}
Chan,~G. K.-L.;\ \ Head-Gordon,~M.  Highly Correlated Calculations with a
  Polynomial Cost Algorithm: {{A}} Study of the Density Matrix Renormalization
  Group,  \textit{J. Chem. Phys.} \textbf{2002,} \textsl{116,} 4462-4476.

\bibitem{Chan_Exact_JCP_2003}
Chan,~G. K.-L.;\ \ Head-Gordon,~M.  Exact Solution (within a Triple-Zeta,
  Double Polarization Basis Set) of the Electronic {{Schr\"odinger}} Equation
  for Water,  \textit{J. Chem. Phys.} \textbf{2003,} \textsl{118,} 8551-8554.

\bibitem{Chan_algorithm_JCP_2004}
Chan,~G. K.-L.  An Algorithm for Large Scale Density Matrix Renormalization
  Group Calculations,  \textit{J. Chem. Phys.} \textbf{2004,} \textsl{120,}
  3172-3178.

\bibitem{Chan_State-of-the-art_JCP_2004}
Chan,~G. K.-L.;\ \ K\'allay,~M.;\ \ Gauss,~J.  State-of-the-Art Density Matrix
  Renormalization Group and Coupled Cluster Theory Studies of the Nitrogen
  Binding Curve,  \textit{J. Chem. Phys.} \textbf{2004,} \textsl{121,}
  6110-6116.

\bibitem{Chan_Density-matrix_JCP_2005}
Chan,~G. K.-L.;\ \ Van~Voorhis,~T.  Density-Matrix Renormalization-Group
  Algorithms with Nonorthogonal Orbitals and Non-{{Hermitian}} Operators, and
  Applications to Polyenes,  \textit{J. Chem. Phys.} \textbf{2005,}
  \textsl{122,} 204101.

\bibitem{Legeza_QC-DMRG_MP_2003}
Legeza,~{\"O}.;\ \ R{\"o}der,~J.;\ \ Hess,~B.~A.  {{QC}}-{{DMRG}} Study of the
  Ionic-Neutral Curve Crossing of {{LiF}},  \textit{Mol. Phys.} \textbf{2003,}
  \textsl{101,} 2019-2028.

\bibitem{Moritz_Convergence_JCP_2005}
Moritz,~G.;\ \ Hess,~B.~A.;\ \ Reiher,~M.  Convergence Behavior of the
  Density-Matrix Renormalization Group Algorithm for Optimized Orbital
  Orderings,  \textit{J. Chem. Phys.} \textbf{2005,} \textsl{122,} 024107.

\bibitem{Moritz_Relativistic_JCP_2005}
Moritz,~G.;\ \ Wolf,~A.;\ \ Reiher,~M.  Relativistic {{DMRG}} Calculations on
  the Curve Crossing of Cesium Hydride,  \textit{J. Chem. Phys.} \textbf{2005,}
  \textsl{123,} 184105.

\bibitem{hachmann2007radical}
Hachmann,~J.;\ \ Dorando,~J.~J.;\ \ Avil{\'e}s,~M.;\ \ Chan,~G. K.-L.  The
  radical character of the acenes: A density matrix renormalization group
  study,  \textit{J. Chem. Phys.} \textbf{2007,} \textsl{127,} 134309.

\bibitem{Zgid_spin_JCP_2008}
Zgid,~D.;\ \ Nooijen,~M.  On the Spin and Symmetry Adaptation of the Density
  Matrix Renormalization Group Method,  \textit{J. Chem. Phys.} \textbf{2008,}
  \textsl{128,} 014107.

\bibitem{Zgid_density_JCP_2008}
Zgid,~D.;\ \ Nooijen,~M.  The Density Matrix Renormalization Group
  Self-Consistent Field Method: {{Orbital}} Optimization with the Density
  Matrix Renormalization Group Method in the Active Space,  \textit{J. Chem.
  Phys.} \textbf{2008,} \textsl{128,} 144116.

\bibitem{Zgid_Obtaining_JCP_2008}
Zgid,~D.;\ \ Nooijen,~M.  Obtaining the Two-Body Density Matrix in the Density
  Matrix Renormalization Group Method,  \textit{J. Chem. Phys.} \textbf{2008,}
  \textsl{128,} 144115.

\bibitem{Kurashige_High-performance_JCP_2009}
Kurashige,~Y.;\ \ Yanai,~T.  High-Performance Ab Initio Density Matrix
  Renormalization Group Method: {{Applicability}} to Large-Scale Multireference
  Problems for Metal Compounds,  \textit{J. Chem. Phys.} \textbf{2009,}
  \textsl{130,} 234114.

\bibitem{zgid2008density}
Zgid,~D.;\ \ Nooijen,~M.  The density matrix renormalization group
  self-consistent field method: Orbital optimization with the density matrix
  renormalization group method in the active space,  \textit{J. Chem. Phys.}
  \textbf{2008,} \textsl{128,} 144116.

\bibitem{ghosh2008orbital}
Ghosh,~D.;\ \ Hachmann,~J.;\ \ Yanai,~T.;\ \ Chan,~G. K.-L.  Orbital
  optimization in the density matrix renormalization group, with applications
  to polyenes and $\beta$-carotene,  \textit{J. Chem. Phys.} \textbf{2008,}
  \textsl{128,} 144117.

\bibitem{sun2017}
Sun,~Q.;\ \ Yang,~J.;\ \ Chan,~G. K.-L.  A General Second Order Complete Active
  Space Self-Consistent-Field Solver for Large-Scale Systems,  \textit{Chem.
  Phys. Lett.} \textbf{2017,} \textsl{683,} 291-299.

\bibitem{wouters2014communication}
Wouters,~S.;\ \ Bogaerts,~T.;\ \ Der~Voort,~P.~V.;\ \ Van~Speybroeck,~V.;\ \
  Van~Neck,~D.  Communication: DMRG-SCF study of the singlet, triplet, and
  quintet states of oxo-Mn(Salen),  \textit{J. Chem. Phys.} \textbf{2014,}
  \textsl{140,} 241103.

\bibitem{Ma_Second-Order_JCTC_2017}
Ma,~Y.;\ \ Knecht,~S.;\ \ Keller,~S.;\ \ Reiher,~M.  Second-{{Order
  Self}}-{{Consistent}}-{{Field Density}}-{{Matrix Renormalization Group}},
  \textit{J. Chem. Theory Comput.} \textbf{2017,} \textsl{13,} 2533-2549.

\bibitem{Ostlund_Thermodynamic_PRL_1995}
\"Ostlund,~S.;\ \ Rommer,~S.  Thermodynamic {{Limit}} of {{Density Matrix
  Renormalization}},  \textit{Phys. Rev. Lett.} \textbf{1995,} \textsl{75,}
  3537-3540.

\bibitem{Rommer_Class_PRB_1997}
Rommer,~S.;\ \ \"Ostlund,~S.  Class of Ansatz Wave Functions for
  One-Dimensional Spin Systems and Their Relation to the Density Matrix
  Renormalization Group,  \textit{Phys. Rev. B} \textbf{1997,} \textsl{55,}
  2164-2181.

\bibitem{Chan_Chapter_ARiCC_2009}
Chan,~G. K.-L.;\ \ Zgid,~D.  Chapter 7 {{The Density Matrix Renormalization
  Group}} in {{Quantum Chemistry}}.   In  \textit{Annual {{Reports}} in
  {{Computational Chemistry}}}, Vol.~5; Wheeler,~R.~A.,\ \ Ed.;  {Elsevier}:
  2009.

\bibitem{mizukami2012more}
Mizukami,~W.;\ \ Kurashige,~Y.;\ \ Yanai,~T.  More $\pi$ electrons make a
  difference: Emergence of many radicals on graphene nanoribbons studied by ab
  initio DMRG theory,  \textit{J. Chem. Theory Comput.} \textbf{2012,}
  \textsl{9,} 401-407.

\bibitem{kurashige2013entangled}
Kurashige,~Y.;\ \ Chan,~G. K.-L.;\ \ Yanai,~T.  Entangled quantum electronic
  wavefunctions of the Mn$_4$CaO$_5$ cluster in photosystem II,  \textit{Nat.
  Chem.} \textbf{2013,} \textsl{5,} 660-666.

\bibitem{Chalupsky_JAmChemSoc_Reactivity_2014}
Chalupsk{\'y},~J.;\ \ Rokob,~T.~A.;\ \ Kurashige,~Y.;\ \ Yanai,~T.;\ \
  Solomon,~E.~I.;\ \ Rul{\'\i}{\v s}ek,~L.;\ \ Srnec,~M.  Reactivity of the
  {{Binuclear Non}}-{{Heme Iron Active Site}} of {{$\Delta^9$ Desaturase
  Studied}} by {{Large}}-{{Scale Multireference Ab Initio Calculations}},
  \textit{J. Am. Chem. Soc.} \textbf{2014,} \textsl{136,} 15977--15991.

\bibitem{sharma2014low}
Sharma,~S.;\ \ Sivalingam,~K.;\ \ Neese,~F.;\ \ Chan,~G. K.-L.  Low-energy
  spectrum of iron--sulfur clusters directly from many-particle quantum
  mechanics,  \textit{Nat. Chem.} \textbf{2014,} \textsl{6,} 927-933.

\bibitem{yana15}
Yanai,~T.;\ \ Kurashige,~Y.;\ \ Mizukami,~W.;\ \ Chalupsky,~J.;\ \ Lan,~T.~N.;\
  \ Saitow,~M.  Density matrix renormalization group for ab initio Calculations
  and associated dynamic correlation methods: A review of theory and
  applications,  \textit{Int. J. Quantum Chem.} \textbf{2015,} \textsl{115,}
  283--299.

\bibitem{freitag2015orbital}
Freitag,~L.;\ \ Knecht,~S.;\ \ Keller,~S.;\ \ Delcey,~M.~G.;\ \ Aquilante,~F.;\
  \ Pedersen,~T.~B.;\ \ Lindh,~R.;\ \ Reiher,~M.;\ \ Gonzalez,~L.  Orbital
  entanglement and CASSCF analysis of the Ru–NO bond in a Ruthenium nitrosyl
  complex,  \textit{Phys. Chem. Chem. Phys.} \textbf{2015,} \textsl{17,}
  14383-14392.

\bibitem{Olivaresamaya2015The}
Olivares-Amaya,~R.;\ \ Hu,~W.;\ \ Nakatani,~N.;\ \ Sharma,~S.;\ \ Yang,~J.;\ \
  Chan,~K.~L.  The ab-initio density matrix renormalization group in practice,
  \textit{J. Chem. Phys.} \textbf{2015,} \textsl{142,} 034102.

\bibitem{Wouters_density_EPJD_2014}
Wouters,~S.;\ \ Van~Neck,~D.  The Density Matrix Renormalization Group for Ab
  Initio Quantum Chemistry,  \textit{Eur. Phys. J. D} \textbf{2014,}
  \textsl{68,} 272.

\bibitem{Pulay_initio_MP_1969}
Pulay,~P.  Ab Initio Calculation of Force Constants and Equilibrium Geometries
  in Polyatomic Molecules,  \textit{Mol. Phys.} \textbf{1969,} \textsl{17,}
  197-204.

\bibitem{Pulay_Direct_AoEST_1977}
Pulay,~P.  Direct {{Use}} of the {{Gradient}} for {{Investigating Molecular
  Energy Surfaces}}.   In  \textit{Applications of {{Electronic Structure
  Theory}}}; Schaefer,~H.~F.,\ \ Ed.;  Modern {{Theoretical Chemistry}}
  {Springer US}: 1977.

\bibitem{Pulay_Systematic_JACS_1979}
Pulay,~P.;\ \ Fogarasi,~G.;\ \ Pang,~F.;\ \ Boggs,~J.~E.  Systematic Ab Initio
  Gradient Calculation of Molecular Geometries, Force Constants, and Dipole
  Moment Derivatives,  \textit{J. Am. Chem. Soc.} \textbf{1979,} \textsl{101,}
  2550-2560.

\bibitem{Pulay_efficient_TCA_1979}
Pulay,~P.  An Efficient Ab Initio Gradient Program,  \textit{Theoret. Chim.
  Acta} \textbf{1979,} \textsl{50,} 299-312.

\bibitem{Iii1984On}
Lengsfield~III,~B.~H.;\ \ Saxe,~P.;\ \ Yarkony,~D.~R.  On the evaluation of
  nonadiabatic coupling matrix elements using SA‐MCSCF/CI wave functions and
  analytic gradient methods. I,  \textit{J. Chem. Phys.} \textbf{1984,}
  \textsl{81,} 4549-4553.

\bibitem{Hellmann}
Hellmann,~H. \textit{Einf\"{u}hrung in die Quantenchemie;} {Leipzig: Deuticke}:
  1937.

\bibitem{Feynman_Forces_PR_1939}
Feynman,~R.~P.  Forces in {{Molecules}},  \textit{Phys. Rev.} \textbf{1939,}
  \textsl{56,} 340-343.

\bibitem{Taylor_Analytical_JCC_1984}
Taylor,~P.~R.  Analytical {{MCSCF}} Energy Gradients: {{Treatment}} of Symmetry
  and {{CASSCF}} Applications to Propadienone,  \textit{J. Comput. Chem.}
  \textbf{1984,} \textsl{5,} 589-597.

\bibitem{Liu2013Multireference}
Liu,~F.;\ \ Kurashige,~Y.;\ \ Yanai,~T.;\ \ Morokuma,~K.  Multireference Ab
  Initio Density Matrix Renormalization Group (DMRG)-CASSCF and DMRG-CASPT2
  Study on the Photochromic Ring Opening of Spiropyran.,  \textit{J. Chem.
  Theory Comput.} \textbf{2013,} \textsl{9,} 4462-4469.

\bibitem{Hu2015Excited}
Hu,~W.;\ \ Chan,~K.~L.  Excited-State Geometry Optimization with the Density
  Matrix Renormalization Group, as Applied to Polyenes,  \textit{J. Chem.
  Theory Comput.} \textbf{2015,} \textsl{11,} 3000-3009.

\bibitem{Nakatani_JChemPhys_Density_2017}
Nakatani,~N.;\ \ Guo,~S.  Density Matrix Renormalization Group ({{DMRG}})
  Method as a Common Tool for Large Active-Space {{CASSCF}}/{{CASPT2}}
  Calculations,  \textit{J. Chem. Phys.} \textbf{2017,} \textsl{146,} 094102.

\bibitem{Ma2017Multiconfigurational}
Ma,~Y.;\ \ Knecht,~S.;\ \ Reiher,~M.  Multiconfigurational Effects in
  Theoretical Resonance Raman Spectra,  \textit{ChemPhysChem} \textbf{2017,}
  \textsl{18,} 384-393.

\bibitem{Yarkony_Electronic_1995}
Yarkony,~D.~R.  Electronic Structure Aspects of Nonadiabatic Processes in
  Polyatomic Systems.   In  \textit{Modern {{Electronic Structure Theory}}};
  Advanced Series in Physical Chemistry {World Scientific Publishing Company}:
  1995.

\bibitem{Schmidt_AnnuRevPhysChem_construction_2003}
Schmidt,~M.~W.;\ \ Gordon,~M.~S.  The Construction and Interpretation of
  {{MCSCF}} Wavefunctions,  \textit{Annu. Rev. Phys. Chem.} \textbf{2003,}
  \textsl{49,} 233--266.

\bibitem{Docken_JChemPhys_LiH_1972}
Docken,~K.~K.;\ \ Hinze,~J.  {{LiH Potential Curves}} and {{Wavefunctions}} for
  {{X}}\,{{$^1\Sigma^+$}}, {{A}}\,{{$^1\Sigma^+$}},
  {{B}}\,{{$^1\Pi$}},\,{{$^1\Sigma^+$}}, and {{$^3\Pi$}},  \textit{J. Chem.
  Phys.} \textbf{1972,} \textsl{57,} 4928--4936.

\bibitem{Werner_JChemPhys_quadratically_1981}
Werner,~H.-J.;\ \ Meyer,~W.  A Quadratically Convergent {{MCSCF}} Method for
  the Simultaneous Optimization of Several States,  \textit{J. Chem. Phys.}
  \textbf{1981,} \textsl{74,} 5794--5801.

\bibitem{Diffenderfer_JPhysChem_Use_1982}
Diffenderfer,~R.~N.;\ \ Yarkony,~D.~R.  Use of the State-Averaged {{MCSCF}}
  Procedure: Application to Radiative Transitions in Magnesium Oxide,
  \textit{J. Phys. Chem.} \textbf{1982,} \textsl{86,} 5098--5105.

\bibitem{Helgaker_Configuration-interaction_TCA_1989}
Helgaker,~T.;\ \ J\o{}rgensen,~P.  Configuration-Interaction Energy Derivatives
  in a Fully Variational Formulation,  \textit{Theor. Chem. Acc.}
  \textbf{1989,} \textsl{75,} 111-127.

\bibitem{Jones2001Analytical}
St{\aa}lring,~J.;\ \ Bernhardsson,~A.;\ \ Lindh,~R.  Analytical gradients of a
  state average MCSCF state and a state average diagnostic,  \textit{Mol.
  Phys.} \textbf{2001,} \textsl{99,} 103-114.

\bibitem{Dupuis_Energy_JCP_1981}
Dupuis,~M.  Energy Derivatives for Configuration Interaction Wave Functions,
  \textit{J. Chem. Phys.} \textbf{1981,} \textsl{74,} 5758-5765.

\bibitem{Osamura_Generalization_JCP_1982}
Osamura,~Y.;\ \ Yamaguchi,~Y.;\ \ Schaefer,~H.~F.  Generalization of Analytic
  Configuration Interaction ({{CI}}) Gradient Techniques for Potential Energy
  Hypersurfaces, Including a Solution to the Coupled Perturbed Hartree-Fock
  Equations for Multiconfiguration {{SCF}} Molecular Wave Functions,
  \textit{J. Chem. Phys.} \textbf{1982,} \textsl{77,} 383-390.

\bibitem{Snyder_atomic_JCP_2015}
Snyder,~J.~W.;\ \ Hohenstein,~E.~G.;\ \ Luehr,~N.;\ \ Mart\'inez,~T.~J.  An
  Atomic Orbital-Based Formulation of Analytical Gradients and Nonadiabatic
  Coupling Vector Elements for the State-Averaged Complete Active Space
  Self-Consistent Field Method on Graphical Processing Units,  \textit{J. Chem.
  Phys.} \textbf{2015,} \textsl{143,} 154107.

\bibitem{Snyder_direct-compatible_JCP_2017}
Snyder,~J.~W.;\ \ Fales,~B.~S.;\ \ Hohenstein,~E.~G.;\ \ Levine,~B.~G.;\ \
  Mart\'inez,~T.~J.  A Direct-Compatible Formulation of the Coupled Perturbed
  Complete Active Space Self-Consistent Field Equations on Graphical Processing
  Units,  \textit{J. Chem. Phys.} \textbf{2017,} \textsl{146,} 174113.

\bibitem{Lengsfield_Nonadiabatic_ACP_2007}
Lengsfield,~B.~H.;\ \ Yarkony,~D.~R.  Nonadiabatic {{Interactions Between
  Potential Energy Surfaces}}: {{Theory}} and {{Applications}},  \textit{Adv.
  Chem. Phys.} \textbf{2007,}  1-71.

\bibitem{wern80}
Werner,~H.-J.;\ \ Meyer,~W.  {A quadratically convergent
  multiconfiguration--self-consistent field method with simultaneous
  optimization of orbitals and CI coefficients},  \textit{J. Chem. Phys.}
  \textbf{1980,} \textsl{73,} 2342--2356.

\bibitem{wern85}
Werner,~H.-J.;\ \ Knowles,~P.~J.  {A second order multiconfiguration SCF
  procedure with optimum convergence},  \textit{J. Chem. Phys.} \textbf{1985,}
  \textsl{82,} 5053--5063.

\bibitem{know85}
Knowles,~P.~J.;\ \ Werner,~H.-J.  {An efficient second-order MC SCF method for
  long configuration expansions},  \textit{Chem. Phys. Lett.} \textbf{1985,}
  \textsl{115,} 259-267.

\bibitem{wern87}
Werner,~H.-J.  {Matrix-formulated direct multiconfiguration self-consistent
  field and multiconfiguration reference configuration-interaction methods},
  \textit{Adv. Chem. Phys.} \textbf{1987,} \textsl{69,} 1-62.

\bibitem{leng81}
Lengsfield~{III},~B.~H.;\ \ Liu,~B.  {A second order MCSCF method for large CI
  expansions},  \textit{J. Chem. Phys.} \textbf{1981,} \textsl{75,} 478-480.

\bibitem{leng82}
Lengsfield~{III},~B.~H.  {General second-order MCSCF theory for large CI
  expansions},  \textit{J. Chem. Phys.} \textbf{1982,} \textsl{77,} 4073-4083.

\bibitem{jorg81}
J{\o}rgensen,~P.;\ \ Olsen,~J.;\ \ Yeager,~D.~L.  {Generalizations of
  Newton-Raphson and multiplicity independent Newton-Raphson approaches in
  multiconfigurational Hartree-Fock theory},  \textit{J. Chem. Phys.}
  \textbf{1981,} \textsl{75,} 5802-5815.

\bibitem{yeag82}
Yeager,~D.~L.;\ \ Lynch,~D.;\ \ Nichols,~J.;\ \ J{\o}rgensen,~P.;\ \ Olsen,~J.
  {Newton-Raphson approaches and generalizations in multiconfigurational
  self-consistent field calculations},  \textit{J. Phys. Chem.} \textbf{1982,}
  \textsl{86,} 2140-2153.

\bibitem{Olsen_Optimization_AiCP_1983}
Olsen,~J.;\ \ Yeager,~D.~L.;\ \ J\o{}rgensen,~P.  Optimization and
  {{Characterization}} of a {{Multiconfigurational Self}}-{{Consistent Field}}
  ({{MCSCF}}) {{State}}.   In  \textit{Advances in {{Chemical Physics}}};
  Prigogine,~I.;\ \ Rice,~S.~A.,\ \ Eds.;  {John Wiley \& Sons, Inc.}: 1983.

\bibitem{Wouters_Thouless_PRB_2013}
Wouters,~S.;\ \ Nakatani,~N.;\ \ Van~Neck,~D.;\ \ Chan,~G. K.-L.  Thouless
  Theorem for Matrix Product States and Subsequent Post Density Matrix
  Renormalization Group Methods,  \textit{Phys. Rev. B} \textbf{2013,}
  \textsl{88,} 075122.

\bibitem{Nakatani_Linear_JCP_2014}
Nakatani,~N.;\ \ Wouters,~S.;\ \ Van~Neck,~D.;\ \ Chan,~G. K.-L.  Linear
  Response Theory for the Density Matrix Renormalization Group: {{Efficient}}
  Algorithms for Strongly Correlated Excited States,  \textit{J. Chem. Phys.}
  \textbf{2014,} \textsl{140,} 024108.

\bibitem{Dorando_Analytic_JCP_2009}
Dorando,~J.~J.;\ \ Hachmann,~J.;\ \ Chan,~G. K.-L.  Analytic Response Theory
  for the Density Matrix Renormalization Group,  \textit{J. Chem. Phys.}
  \textbf{2009,} \textsl{130,} 184111.

\bibitem{Haegeman_Time-Dependent_PRL_2011}
Haegeman,~J.;\ \ Cirac,~J.~I.;\ \ Osborne,~T.~J.;\ \ Pi{\v z}orn,~I.;\ \
  Verschelde,~H.;\ \ Verstraete,~F.  Time-{{Dependent Variational Principle}}
  for {{Quantum Lattices}},  \textit{Phys. Rev. Lett.} \textbf{2011,}
  \textsl{107,} 070601.

\bibitem{Vanderstraeten_Tangent-space_SPLN_2019}
Vanderstraeten,~L.;\ \ Haegeman,~J.;\ \ Verstraete,~F.  Tangent-Space Methods
  for Uniform Matrix Product States,  \textit{SciPost Phys. Lect. Notes}
  \textbf{2019,}  7.

\bibitem{roos1980complete}
Roos,~B.~O.;\ \ Taylor,~P.~R.;\ \ Siegbahn,~P.~E.  A complete active space SCF
  method (CASSCF) using a density matrix formulated super-CI approach,
  \textit{Chem. Phys.} \textbf{1980,} \textsl{48,} 157-173.

\bibitem{szalay2011multiconfiguration}
Szalay,~P.~G.;\ \ M\"uller,~T.;\ \ Gidofalvi,~G.;\ \ Lischka,~H.;\ \
  Shepard,~R.  Multiconfiguration self-consistent field and multireference
  configuration interaction methods and applications,  \textit{Chem. Rev.}
  \textbf{2011,} \textsl{112,} 108-181.

\bibitem{Schollwock_density-matrix_PTRSA_2011}
Schollw\"ock,~U.  The Density-Matrix Renormalization Group: A Short
  Introduction,  \textit{Phil. Trans. R. Soc. A} \textbf{2011,} \textsl{369,}
  2643-2661.

\bibitem{kell15a}
Keller,~S.;\ \ Dolfi,~M.;\ \ Troyer,~M.;\ \ Reiher,~M.  An efficient matrix
  product operator representation of the quantum-chemical Hamiltonian,
  \textit{J. Chem. Phys.} \textbf{2015,} \textsl{143,} 244118.

\bibitem{Dolgov2014_BlockEigenvalues}
Dolgov,~S.~V.;\ \ Khoromskij,~B.~N.;\ \ Oseledets,~I.~V.;\ \ Savostyanov,~D.~V.
   {Computation of extreme eigenvalues in higher dimensions using block tensor
  train format},  \textit{Comput. Phys. Commun.} \textbf{2014,} \textsl{185,}
  1207--1216.

\bibitem{Bernhardsson_MolPhys_direct_1999}
Bernhardsson,~A.;\ \ Lindh,~R.;\ \ Olsen,~J.;\ \ F\"{u}lscher,~M.  A Direct
  Implementation of the Second-Order Derivatives of Multiconfigurational
  {{SCF}} Energies and an Analysis of the Preconditioning in the Associated
  Response Equation,  \textit{Mol. Phys.} \textbf{1999,} \textsl{96,} 617--628.

\bibitem{OpenMOLCAS}
Fdez.~Galv{\'a}n,~I.; Vacher,~M.; Alavi,~A.; Angeli,~C.; Aquilante,~F.;
  Autschbach,~J.; Bao,~J.~J.; Bokarev,~S.~I.; Bogdanov,~N.~A.; Carlson,~R.~K.;
  Chibotaru,~L.~F.; Creutzberg,~J.; Dattani,~N.; Delcey,~M.~G.; Dong,~S.~S.;
  Dreuw,~A.; Freitag,~L.; Frutos,~L.~M.; Gagliardi,~L.; Gendron,~F.;
  Giussani,~A.; Gonz{\'a}lez,~L.; Grell,~G.; Guo,~M.; Hoyer,~C.~E.;
  Johansson,~M.; Keller,~S.; Knecht,~S.; Kova{\u{c}}evi{\'c},~G.;
  K{\"a}llman,~E.; Li~Manni,~G.; Lundberg,~M.; Ma,~Y.; Mai,~S.; Malhado,~J.~P.;
  Malmqvist,~P.~{\AA}.; Marquetand,~P.; Mewes,~S.~A.; Norell,~J.; Olivucci,~M.;
  Oppel,~M.; Phung,~Q.~M.; Pierloot,~K.; Plasser,~F.; Reiher,~M.; Sand,~A.~M.;
  Schapiro,~I.; Sharma,~P.; Stein,~C.~J.; S{\o}rensen,~L.~K.; Truhlar,~D.~G.;
  Ugandi,~M.; Ungur,~L.; Valentini,~A.; Vancoillie,~S.; Veryazov,~V.;
  Weser,~O.; Weso{\l}owski,~T.~A.; Widmark,~P.-O.; Wouters,~S.; Zech,~A.;
  Zobel,~J.~P.; Lindh,~R. OpenMolcas: From Source Code to Insight. \textit{J.
  Chem. Theory Comput.} \textbf{2019}, \textsl{15}, 5925-5964.

\bibitem{helgaker2014molecular}
Helgaker,~T.;\ \ J{\o}rgensen,~P.;\ \ Olsen,~J. \textit{Molecular
  electronic-structure theory;} John Wiley \& Sons: 2014.

\bibitem{Fdez.Galvan_Analytical_JCTC_2016}
Fdez.~Galv\'an,~I.;\ \ Delcey,~M.~G.;\ \ Pedersen,~T.~B.;\ \ Aquilante,~F.;\ \
  Lindh,~R.  Analytical {{State}}-{{Average Complete}}-{{Active}}-{{Space
  Self}}-{{Consistent Field Nonadiabatic Coupling Vectors}}: {{Implementation}}
  with {{Density}}-{{Fitted Two}}-{{Electron Integrals}} and {{Application}} to
  {{Conical Intersections}},  \textit{J. Chem. Theory Comput.} \textbf{2016,}
  \textsl{12,} 3636-3653.

\bibitem{kell16}
Keller,~S.;\ \ Reiher,~M.  {Spin-adapted matrix product states and operators},
  \textit{J. Chem. Phys.} \textbf{2016,} \textsl{144,} 134101.

\bibitem{dunn89}
Dunning~Jr,~T.~H.  {Gaussian basis sets for use in correlated molecular
  calculations. I. The atoms boron through neon and hydrogen},  \textit{J.
  Chem. Phys.} \textbf{1989,} \textsl{90,} 1007-1023.

\bibitem{Roos_New_JPCA_2005}
Roos,~B.~O.;\ \ Lindh,~R.;\ \ Malmqvist,~P.-A.;\ \ Veryazov,~V.;\ \
  Widmark,~P.-O.  New {{Relativistic ANO Basis Sets}} for {{Transition Metal
  Atoms}},  \textit{J. Phys. Chem. A} \textbf{2005,} \textsl{109,} 6575-6579.

\bibitem{Liu_Theoretical_JACS_2009}
Liu,~F.;\ \ Liu,~Y.;\ \ De~Vico,~L.;\ \ Lindh,~R.  Theoretical {{Study}} of the
  {{Chemiluminescent Decomposition}} of {{Dioxetanone}},  \textit{J. Am. Chem.
  Soc.} \textbf{2009,} \textsl{131,} 6181-6188.

\bibitem{Carpenter_Heavy-atom_JACS_1983}
Carpenter,~B.~K.  Heavy-Atom Tunneling as the Dominant Pathway in a
  Solution-Phase Reaction? {{Bond}} Shift in Antiaromatic Annulenes,
  \textit{J. Am. Chem. Soc.} \textbf{1983,} \textsl{105,} 1700-1701.

\bibitem{Carsky_Heavy-atom_TCA_1992}
{\v C}\'arsky,~P.;\ \ Michl,~J.  Heavy-Atom Tunneling in Cyclobutadiene:{{Ab}}
  Initio Calculation of the Intensities {{ofagRaman}} Lines,  \textit{Theoret.
  Chim. Acta} \textbf{1992,} \textsl{84,} 125-133.

\bibitem{Schoonmaker_Quantum_JCP_2018}
Schoonmaker,~R.;\ \ Lancaster,~T.;\ \ Clark,~S.~J.  Quantum Mechanical
  Tunneling in the Automerization of Cyclobutadiene,  \textit{J. Chem. Phys.}
  \textbf{2018,} \textsl{148,} 104109.

\bibitem{Eckert-Maksic_Automerization_JCP_2006}
Eckert-Maksi\'c,~M.;\ \ Vazdar,~M.;\ \ Barbatti,~M.;\ \ Lischka,~H.;\ \
  Maksi\'c,~Z.~B.  Automerization Reaction of Cyclobutadiene and Its Barrier
  Height: {{An}} Ab Initio Benchmark Multireference Average-Quadratic Coupled
  Cluster Study,  \textit{J. Chem. Phys.} \textbf{2006,} \textsl{125,} 064310.

\bibitem{Schmidt_Kinetics_JACS_1978}
Schmidt,~S.~P.;\ \ Schuster,~G.~B.  Kinetics of Unimolecular Dioxetanone
  Chemiluminescence. {{Competitive}} Parallel Reaction Paths,  \textit{J. Am.
  Chem. Soc.} \textbf{1978,} \textsl{100,} 5559-5561.

\bibitem{Adam_Cyclic_JACS_1979}
Adam,~W.;\ \ Cueto,~O.  Cyclic Peroxides. 81. {{Fluorescer}}-Enhanced
  Chemiluminescence of .Alpha.-Peroxylactones via Electron Exchange,
  \textit{J. Am. Chem. Soc.} \textbf{1979,} \textsl{101,} 6511-6515.

\bibitem{Turro_Chemiluminescent_JACS_1980}
Turro,~N.~J.;\ \ Chow,~M.-F.  Chemiluminescent Thermolysis of
  Alpha-Peroxylactones,  \textit{J. Am. Chem. Soc.} \textbf{1980,}
  \textsl{102,} 5058-5064.

\bibitem{Liu_CASSCFCASPT2_CPL_2009}
Liu,~F.;\ \ Liu,~Y.;\ \ Vico,~L.~D.;\ \ Lindh,~R.  A {{CASSCF}}/{{CASPT2}}
  Approach to the Decomposition of Thiazole-Substituted Dioxetanone:
  {{Substitution}} Effects and Charge-Transfer Induced Electron Excitation,
  \textit{Chem. Phys. Lett.} \textbf{2009,} \textsl{484,} 69-75.

\bibitem{Navizet_Chemistry_C_2011}
Navizet,~I.;\ \ Liu,~Y.-J.;\ \ Ferr\'e,~N.;\ \ Roca-Sanju\'an,~D.;\ \ Lindh,~R.
   The {{Chemistry}} of {{Bioluminescence}}: {{An Analysis}} of {{Chemical
  Functionalities}},   \textbf{2011,} \textsl{12,} 3064-3076.

\bibitem{Yue_Theoretical_JPCL_2015}
Yue,~L.;\ \ Lan,~Z.;\ \ Liu,~Y.-J.  The {{Theoretical Estimation}} of the
  {{Bioluminescent Efficiency}} of the {{Firefly}} via a {{Nonadiabatic
  Molecular Dynamics Simulation}},  \textit{J. Phys. Chem. Lett.}
  \textbf{2015,} \textsl{6,} 540-548.

\bibitem{Farahani_combined_RA_2017}
Farahani,~P.;\ \ Oliveira,~M.~A.;\ \ Fdez.~Galv\'an,~I.;\ \ Baader,~W.~J.  A
  Combined Theoretical and Experimental Study on the Mechanism of
  Spiro-Adamantyl-1,2-Dioxetanone Decomposition,  \textit{RSC Adv.}
  \textbf{2017,} \textsl{7,} 17462-17472.

\bibitem{Vacher_Chemi-_CR_2018}
Vacher,~M.;\ \ Fdez.~Galv\'an,~I.;\ \ Ding,~B.-W.;\ \ Schramm,~S.;\ \
  Berraud-Pache,~R.;\ \ Naumov,~P.;\ \ Ferr\'e,~N.;\ \ Liu,~Y.-J.;\ \
  Navizet,~I.;\ \ Roca-Sanju\'an,~D.;\ \ Baader,~W.~J.;\ \ Lindh,~R.  Chemi-
  and {{Bioluminescence}} of {{Cyclic Peroxides}},  \textit{Chem. Rev.}
  \textbf{2018,} \textsl{118,} 6927-6974.

\bibitem{Yue_Two_JCTC_2019}
Yue,~L.;\ \ Liu,~Y.-J.  Two {{Conical Intersections Control Luminol
  Chemiluminescence}},  \textit{J. Chem. Theory Comput.} \textbf{2019,}
  \textsl{15,} 1798-1805.

\end{thebibliography}
\providecommand{\refin}[1]{\\ \textbf{Referenced in:} #1}

%\newpage
%\vspace*{0.9cm}
%\par
%\textbf{Table-of-Contents Graphic}
%\par
%\vspace*{0.9cm}
%\includegraphics[height=4cm]{toc.pdf}

\end{document}